\documentclass[fleqn,usenatbib]{mnras}

\usepackage{newtxtext,newtxmath}
\usepackage[T1]{fontenc}
\usepackage{graphicx}	% Including figure files
\usepackage{amsmath}	% Advanced maths commands
\usepackage{psfrag}
\usepackage{bm}
\usepackage{multirow}
\usepackage{ulem}
\def\lsim{~\rlap{$<$}{\lower 1.0ex\hbox{$\sim$}}}

\def\gsim{~\rlap{$>$}{\lower 1.0ex\hbox{$\sim$}}}

\def\HI{\ion{H}{I}~}
\def\V{\mathcal{V}}
\def\u{{\bf U}} 
\def\kv{{\mathbf k}}
\def\cl{{C_{\ell}(\Delta\nu)}}

\defcitealias{Bh18}{Paper~I} 
\defcitealias{Pal20}{Paper~II} 
\defcitealias{Ch21}{Ch21} 

\title[Intensity Mapping at $z=2.28$ with uGMRT]{Towards 21-cm Intensity Mapping at $z=2.28$ with uGMRT using the Tapered Gridded Estimator I: Foreground Avoidance}

\author[Pal S et al.]{
Srijita Pal,$^{1}$\thanks{E-mail:srijitapal.phy@gmail.com}
Kh. Md. Asif Elahi,$^{1}$
Somnath Bharadwaj,$^{1}$\thanks{E-mail: somnath@phy.iitkgp.ac.in}
Sk. Saiyad Ali,$^{2}$
Samir Choudhuri,$^{3,4}$
\newauthor
Abhik Ghosh,$^{5}$
Arnab Chakraborty,$^{6}$
Abhirup Datta,$^{7}$
Nirupam Roy,$^{8}$
Madhurima Choudhury,$^{9,7}$
\newauthor
Prasun Dutta$^{10}$
\\
$^{1}$ Department of Physics and Centre for Theoretical Studies, IIT Kharagpur, Kharagpur 721 302, India\\
$^{2}$ Department of Physics, Jadavpur University, Kolkata 700032, India\\
$^{3}$ Astronomy Unit, Queen Mary University of London, Mile End Road, London E1 4NS, United Kingdom\\
$^{4}$ Centre for Strings, Gravitation and Cosmology, Department of Physics, Indian Institute of Technology Madras, Chennai 600036, India\\
$^{5}$ Department of Physics, Banwarilal Bhalotia College, Asansol, West Bengal-713303, India\\
$^{6}$ Department of Physics and McGill Space Institute, McGill University, Montreal, QC, Canada H3A 2T8\\
$^{7}$ Discipline of Astronomy, Astrophysics and Space Engineering, Indian Institute of Technology Indore, Indore 453552, India\\
$^{8}$ Department of Physics, Indian Institute of Science, Bangalore 560012, India\\
$^{9}$ ARCO (Astrophysics Research Center), Department of Natural Sciences, The Open University of Israel, 1 University Road, PO Box 808, Ra’anana 4353701, Israel\\
$^{10}$ Department of Physics, IIT (BHU), Varanasi, 221005 India
}

\date{Accepted XXX. Received YYY; in original form ZZZ}

\pubyear{2022}

\begin{document}
\label{firstpage}
\pagerange{\pageref{firstpage}--\pageref{lastpage}}
\maketitle

\begin{abstract}
The post-reionization $(z \le 6)$ neutral hydrogen (\HI) 21-cm intensity mapping signal holds the potential to probe the large scale structures, study the expansion history and constrain various cosmological parameters. Here we apply the Tapered Gridded Estimator (TGE) to estimate $P(k_{\perp},k_{\parallel})$ the power spectrum of  the $z = 2.28$ $(432.8\, {\rm  MHz})$ redshifted 21-cm  signal using a $24.4\,{\rm MHz}$ sub-band  drawn from uGMRT Band 3 observations of   European Large-Area ISO Survey-North 1 (ELAIS-N1). The TGE allows us to taper the sky response which suppresses the foreground contribution from sources in the periphery of the telescope's field of view. We apply the TGE on the measured visibility data to estimate the multi-frequency angular power spectrum (MAPS) $C_{\ell}(\Delta\nu)$ from which we determine $P(k_{\perp},k_{\parallel})$  using maximum-likelihood which naturally overcomes the issue of missing frequency channels  (55 \%  here). The entire methodology is validated using simulations. For the data, using the foreground avoidance technique, we obtain  a $2\,\sigma$ upper limit of $\Delta^2(k) \le (133.97)^2 \, {\rm mK}^{2}$ for the 21-cm brightness temperature fluctuation at $k = 0.347 \, \textrm{Mpc}^{-1}$.  This corresponds to $[\Omega_{\rm \HI}b_{\rm \HI}] \le 0.23$, where $\Omega_{\rm \HI}$ and $b_{\rm \HI}$ respectively denote the cosmic \HI mass density  and the \HI  bias parameter. A previous work has analyzed $8 \, {\rm MHz}$ of the same data at $z=2.19$, and reported $\Delta^{2}(k) \le (61.49)^{2} \, {\rm mK}^{2}$ and $[\Omega_{\rm \HI} b_{\rm \HI}] \le 0.11$ at $k=1 \, {\rm Mpc}^{-1}$. The upper limits presented here are still orders of magnitude larger than the expected signal corresponding to $\Omega_{\rm \HI} \sim 10^{-3}$ and $b_{\rm \HI} \sim 2 $.
\end{abstract}


%%%%%%%%%%%%%%%%%%%%%%%%%%%%%%%%%%%%%%%%%%%%%%%%%%

%%%%%%%%%%%%%%%%% BODY OF PAPER %%%%%%%%%%%%%%%%%%

\section{Introduction}
\label{s1}

Intensity mapping with the neutral hydrogen (\HI) 21-cm line is a progressing tool to probe the  large scale structures of the post-reionization Universe \citep{BNS, BS01}. It can independently assess the expansion history of the Universe by measuring the Baryon Acoustic Oscillation (BAO) in the 21-cm power spectrum (PS) \citep{w08, Chang08, Seo_2010}. It is also possible to constrain various cosmological parameters  using measurements of the 21-cm PS without reference to the BAO \citep{Bh09, Visbal_2009}.  Further it is possible to quantify higher order statistics, such as bispectrum, to study non-Gaussianity \citep{BA5, Hazra2012}. 

There has been several successful 21-cm intensity mapping experiments \citep{Pen2009a, chang10, masui2013, SW13, Anderson2018, Wolz2021} at  low-redshifts $(z<1)$. Most of these experiments have used single dish telescopes, and they have detected the 21-cm signal by cross-correlating their measurements with existing galaxy redshift surveys. \citet{SW13} have detected the 21-cm intensity mapping signal in auto-correlation at $z=0.8$ using the Green Bank Telescope. Recently, CHIME\footnote{\url{https://chime-experiment.ca/en/}} \citep{chime22} has made the first interferometric 21-cm intensity mapping measurements in the redshift range $0.78<z<1.43$ using cross-correlations with luminous red galaxies, emission line galaxies, and quasars. Several other radio interferometers such as BINGO\footnote{\url{https://bingotelescope.org/}} \citep{Wuensche_2019}, the Tianlai project\footnote{\url{http://tianlai.bao.ac.cn/}} \citep{tian},  HIRAX\footnote{\url{https://hirax.ukzn.ac.za/}} \citep{Newburgh16} and MeerKAT\footnote{\url{https://www.sarao.ac.za/science/meerkat/}} \citep{Kennedy21} particularly focus on measuring the BAO to study the nature of Dark Energy. The Ooty Radio Telescope (ORT; \citealt{GS71}) is being upgraded to the Ooty Wide Field Array (OWFA\footnote{\url{http://rac.ncra.tifr.res.in/ort.html}}; \citealt{OWFA}) to measure the $21$-cm PS at $z\sim3.35$. Furthermore, the next-generation intensity mapping surveys, with the Square Kilometer Array (SKA\footnote{\url{https://www.skatelescope.org/}}; \citealt{SKA15}) hold the potential to provide a large cosmological window to  the post-reionization era.

The Giant Metrewave Radio Telescope (GMRT\footnote{\url{http://www.gmrt.ncra.tifr.res.in/}};  \citealt{swarup91}) is sensitive to the post-reionization 21-cm intensity mapping signal from a broad redshift range  ($z \le 6$; \citealt{Bharadwaj-Pandey-2003, bh_sri2004}). In an  effort towards this  \cite{ghosh1, ghosh2} have analyzed  $610 \, {\rm MHz}$ GMRT data  to place an upper limit $[\bar{x}_{\rm \HI} \, b_{\rm \HI}] < 2.9$ where  $\bar{x}_{\rm \HI}$ and $ b_{\rm \HI}$ are  the mean neutral fraction and bias parameter respectively at redshift $z=1.32$. We note that  this corresponds to $[\Omega_{\rm \HI} b_{\rm \HI}] < 0.11$ where $\Omega_{\rm \HI}$ is the comoving \HI mass density in units of the present critical density. In a recent work  \cite{Ch21} (hereafter Ch21) have analyzed several $8 \, {\rm MHz} $ subsets drawn from  $200 \, \rm{MHz}$ upgraded GMRT (uGMRT; \citealt{uGMRT}) data to estimate the 21-cm PS at multiple redshifts in the range $1.96<z<3.58$ and place  the  upper limits  $[\Omega_{\rm \HI} b_{\rm \HI}] < 0.09, 0.11,0.12, \, {\rm and} \, 0.24$ at $z=1.96,2.19,2.62 \, {\rm and} \, 3.58$ respectively. 
 
The Tapered Gridded Estimator (TGE; \citealt{samir14, samir16}) is a visibility based PS estimator which allows us to taper the sky response to  suppress the contribution from bright sources located in the side-lobes and the periphery of the telescope's field of view. Further, the TGE works with the  gridded visibilities which makes it computationally fast.  The TGE also uses the measured visibility  data to  internally evaluate the noise bias  and subtracts this to provide an unbiased estimate of the PS. \cite{Bh18} (hereafter, Paper I) have proposed the TGE to estimate  $C_{\ell}(\Delta\nu)$ the multi-frequency angular power spectrum (MAPS; \citealt{KD07, Mondal19}) which   characterizes the second order statistics of the sky signal jointly as a function of the angular multipole $\ell$ and frequency separation $\Delta\nu$. They use  $C_{\ell}(\Delta\nu)$ to determine the cylindrical power spectrum  of the 21-cm brightness temperature fluctuations $P(k_{\perp},k_{\parallel})$ which  is related to  $C_{\ell}(\Delta \nu)$ through a Fourier transform with respect to $\Delta \nu$.  Using simulated visibility data,  they show that this estimator can accurately recover  the input model PS even when  $80\%$  randomly chosen frequency channels are flagged. A salient feature of this estimator is that it only uses the available data, and it is not necessary to make any assumption regarding the data values in the missing frequency channels.  In a recent paper \citet{Pal20} (hereafter Paper II)  have demonstrated the capabilities of the TGE by using the TGE to estimate $P(k_{\perp},k_{\parallel})$ from a 150 MHz GMRT observational data where $47 \%$ of the frequency channels are flagged due to Radio Frequency Interference (RFI).  They  obtain a $ 2 \sigma$  upper limit of $(72.66)^2\,{\rm K}^{2}$ on the mean squared \HI 21-cm brightness temperature fluctuations at $k = 1.59\, {\rm Mpc}^{-1}$. We note that the two dimensional (2D) TGE for the angular power spectrum $C_{\ell}$ has been extensively used  to study the foregrounds for cosmological 21-cm observations \citep{samir17a, samir20, Cha1, M20}  and also magnetohydrodynamics turbulence in supernova remnants \citep{Preetha19, Preetha21}.

In this work we consider uGMRT Band 3 $(300-500\,{\rm MHz})$ data  of the ELAIS N1  ﬁeld. \cite{Cha2}  have analysed this data and used the 2D TGE to study the angular and spectral variation of $C_{\ell}(\nu)$ for the diffuse galactic synchrotron emission. As mentioned earlier, \citetalias{Ch21} have  analysed this data using a  delay spectrum approach to estimate the PS of the 21-cm intensity mapping signal. The difficulty arises because the missing frequency channels (flagged due to RFI) introduce artefacts in delay space  which corrupt the estimated PS.   \citetalias{Ch21} have overcome this by using  one dimensional complex CLEAN \citep{Parsons_2009}  to compensate  for the missing frequency channels. Considering   the same data, the present work  uses a bandwidth of $24.4\,{\rm MHz}$ centred at $432.8\,{\rm MHz}$. Here we have applied the  TGE to estimate the MAPS and PS $P(k_{\perp},k_{\parallel})$. We study the capabilities of this  estimator to (1) suppress the wide-field foregrounds, and (2) deal with the missing frequency channels.  We also present results for the spherically binned power spectrum, and present an upper limit for  $[\Omega_{\rm \HI} b_{\rm \HI}]$ at $z=2.28$.

The  paper is arranged as follows. Section~\ref{s2} summarizes  the observations  and the preliminary processing of the data which has been used here. Section~\ref{s3} presents the methodology how TGE is used to estimate $C_{\ell}(\Delta \nu)$ and $P(k_{\perp}, k_{\parallel})$ from the observed visibility data, and  in Section~\ref{sec:simulation} we have validated the methodology using simulations. Our results are presented in Section~\ref{s6}, and we present a summary and conclusion in Section~\ref{s7}.

Throughout this paper,  unless mentioned otherwise, we have used  a $\Lambda$CDM cosmology with the  parameters $\Omega_{m} = 0.309$,  $h = 0.67$, $n_s = 0.965$, and $\Omega_{b}h^2 = 0.0224$ which are in  reasonable agreement with the present observations \citep{Planck18f}.   

\section{Observations and data analysis}
\label{s2}

\begin{table}
    \centering
    \caption{Observation summary} 
    \label{t_1}
    \begin{tabular}{|l|c|}
        \hline
        \hline
        Working antennas & $28$\\
        \hline
        Central Frequency & $400$ MHz  \\
        \hline
        Number of Channels & $8192$ \\
        \hline
        Channel width & $24.4$ kHz \\
        \hline
        Bandwidth  & $200$ MHz \\
        \hline
        Total observation time &  $25 $ h \\
        \hline
        Integration time & $2$ s \\
        \hline
        Target field  $(\alpha,\delta)_{2000}$ &  ($16^{h}10^{m}1^{s}$,\\ 
         & $+54^{\circ}30^{'}36^{''}$) \\
        \hline
        Galactic coordinates $(l,b)$ & $86.95^{\circ},+44.48^{\circ}$ \\
        \hline
        \hline
    \end{tabular}
\end{table}

The observations were carried out using the GMRT array configuration \citep{swarup91}. The recently upgraded version of GMRT (uGMRT) provides a frequency coverage of $120-1500\,\textrm{MHz}$ with $400~ \textrm{MHz}$ maximum instantaneous available bandwidth and an improved receiver system with higher $G/T_{sys}$ for high dynamic range imaging \citep{uGMRT}. 
We have observed the field European Large-Area ISO Survey-North 1 (ELAIS-N1; $\alpha_{2000}=16^{h}10^{m}1^{s}, \delta_{2000}=54^{\circ}30^{'}36^{''}$) with the uGMRT at Band 3 ($300-500\,{\rm MHz}$) during May 2017 for 25 hours over four days. The primary calibrators 3C286  and  3C48 have been used to scale the overall flux of the observation. We have also used a nearby phase calibrator (J1549+506) to correct the antenna gains' temporal variation. The total bandwidth of the observation is 200 MHz with a frequency resolution of 24.4 kHz. We have taken the data with a high time resolution (2 s) to identify and remove the RFI. The observations were carried out mainly at night to minimize the RFI. The relevant observational parameters have been summarized in Table \ref{t_1}.

\begin{figure}
    \centering
    \includegraphics[width=\columnwidth]{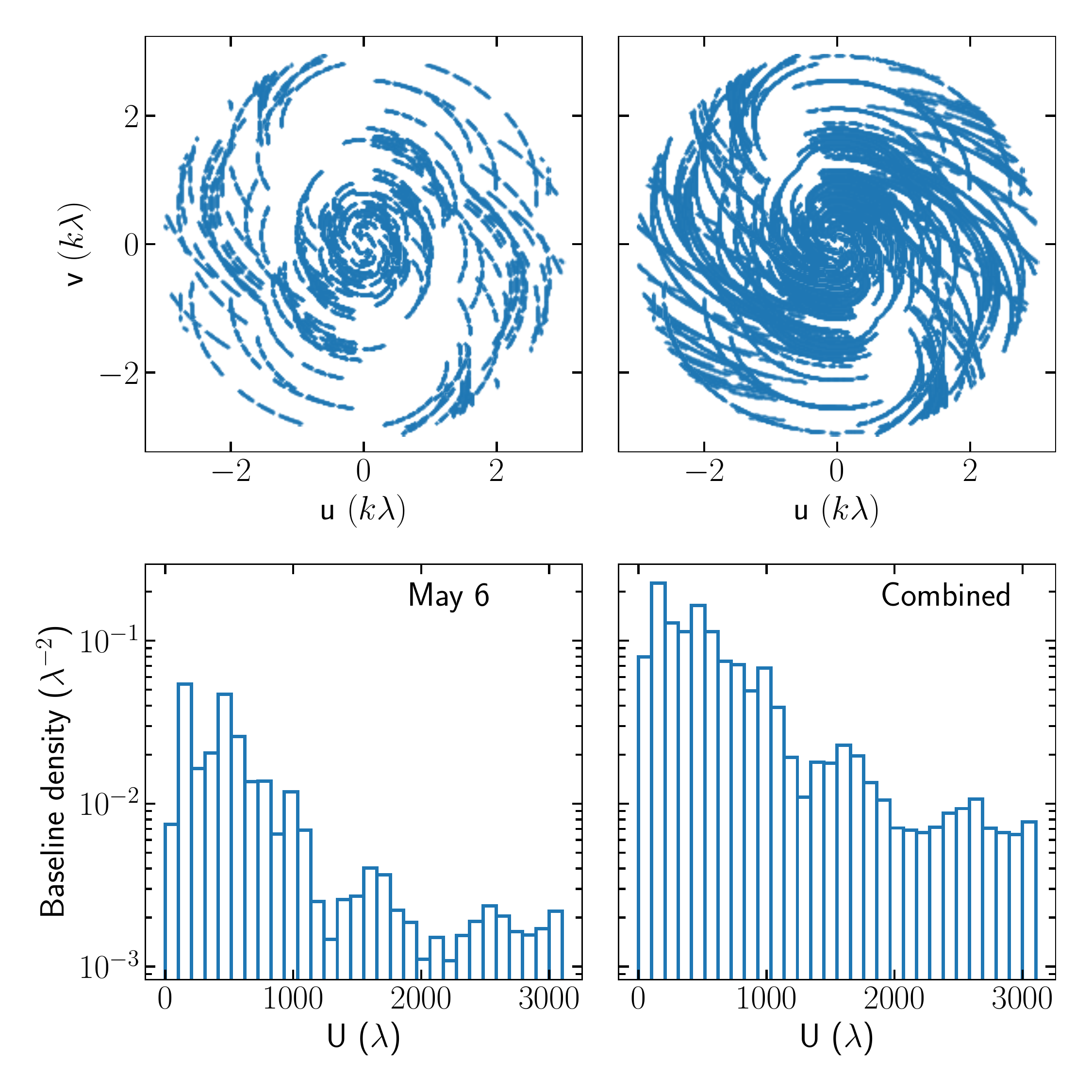}
    \caption{The upper panels show the uv-coverage for the May 6 data (left panel) and the combined  nights data (right panel) considering  baselines of length  $U \leq 3000 \lambda$ at $\nu_{c}=432.8\,{\rm MHz}$. The corresponding baseline density (number of baselines per unit area of the uv plane) is shown as a function of $U$  in the lower panels.}
    \label{fig:uv}
\end{figure}

The details of the initial data analysis are given in \cite{Cha1, Cha2}. Here, we briefly summarize the flagging and calibration steps adopted. For initial RFI flagging, we use the {\scriptsize AOFLAGGER} which detects any anomaly in time-frequency domain per baseline per polarization and discards the corrupt data \citep{Off10, Off12}. We discard $5\%$ of the total number of channels  ($2.5\% $ on each side) due to a bad bandshape at the edge of the bandwidth. We take the direction-independent approach to calibrate the data using the Common Astronomy Software Applications ({\scriptsize CASA}; \citealt{casa07}). We start with an initial gain and bandpass calibration of the primary calibrators and remove residual RFIs from the calibrated data using the {\scriptsize RFLAG} routine of {\scriptsize CASA}. After doing this twice, we perform a final bandpass and delay calibration on the primary calibrators. We next rectify the temporal variation of the amplitude and the phase of the antenna gains using the secondary calibrator. Finally, we apply the gain solutions to the target field and split this out for imaging. The final off-source r.m.s. noise is  $\sim 15\mu  \textrm{Jy}\,\textrm{beam}^{-1}$ which is nearly $3.5$ times higher than the theoretically expected noise. Note that we have calculated the theoretical noise using the specifications of uGMRT for Band 3 \citep{uGMRT}, and considering the parameters of our observations. The excess noise in the image can thus be caused due to a difference in the system temperature depending on the direction of sky we have observed; this may also be caused due to residual deconvolution errors during imaging, residual calibration errors etc. Hence, we expect this excess noise to further contribute as excess power in the estimated power spectrum as residual foregrounds and systematics, as well as increase the error budget (the r.m.s. fluctuations) of the power spectrum estimated from the data \citep{JK20,JK22}. Post four rounds of self-calibration (only phase), we have identified and subtracted out the compact and discrete sources with flux densities $> 100\mu  \textrm{Jy}$ within  an area of $1.8 \, {\rm deg^2}$ using the task {\scriptsize UVSUB} in {\scriptsize CASA}.  The residual visibility data is used for the subsequent  analysis presented here. 

The baseline distribution and various telescope parameters change with frequency across the $200 \, {\rm MHz}$ uGMRT bandwidth. For the subsequent  analysis  we have divided the total  bandwidth into eight  sub-bands (namely, sub-bands 0 to  7). The bandwidth of each sub-band $\sim 24.4 \, {\rm MHz}$ is less than $10 \%$ of the central frequency of the corresponding sub-band, which  allows us to ignore the baseline migration and change in telescope parameters within the sub-band. This is a fair assumption considering Section \ref{sec:simulation} where we validate our estimator for the data considered here. We have checked the quality of the data in each of the sub-bands in terms of the percentage of flagged channels and the variance of the data. The May 6 observation from the sub-band 2 is found to have the least percentage of flagging and the smallest visibility r.m.s. Guided by this,  we have entirely restricted the subsequent analysis of this paper to sub-band 2 which is centred at $\nu_{c}=432.8\,{\rm MHz}$ and contains 1000 channels with spectral resolution $\Delta\nu_{c}=24.4\,\,{\rm kHz}$. In addition to the individual nights data, we have also combined the four nights  data  using the {\scriptsize CONCAT} task of {\scriptsize CASA}. The flagging statistics and r.m.s. values of the individual nights and the combined nights data are given in Table \ref{tab:flagandrms}. 

The upper panels of Figure~\ref{fig:uv}  show the baseline distributions for the  May 6 data (left panel) and the combined nights data (right panel) for a single channel at the central frequency.  The lower panels show the baseline density, {\it i.e.}, the number of baselines per unit area of the uv-plane, as a function of baseline length $U=\lvert{\bf U}\rvert$. 

\begin{table}
\centering
\caption{Considering the sub-band 2 with central frequency $\nu_{c}=432.8\,{\rm MHz}$, spectral resolution $\Delta\nu_{c}=24.4\,\,{\rm kHz}$ and bandwidth $B_{bw}=24.4\,\,{\rm MHz}$, we tabulate the flagging fraction and r.m.s. of the visibilities $\sigma_{N}$ for different nights of observation}
\label{tab:flagandrms}
\begin{tabular}{|c|c|c|}
\hline
\hline
Night of observation & flag (\%) & r.m.s. $\sigma_{N}$ (Jy)\\
\hline
\hline
May 5 & 70.97 & 0.431337 \\
\hline
May 6 & 13.07 & 0.394103 \\
\hline
May 7 & 42.75 & 0.473156 \\
\hline
May 27 & 71.29 & 0.445912 \\
\hline
All nights combined & 54.81 & 0.430112 \\
\hline
\hline
\end{tabular}
\end{table}

\section{Methodology}
\label{s3}

The multi-frequency angular power spectrum (MAPS) $C_{\ell}(\nu_a, \nu_b)$ jointly characterizes the statistical properties of the sky signal as a function of the angular multipoles and frequency. The brightness temperature fluctuations in the sky is decomposed in terms of the spherical harmonics $Y_{\ell}^{\rm m}(\hat{\bm{n}})$ as,
\begin{equation}
\delta T_{\rm b} (\hat{\bm{n}},\,\nu)=\sum_{\ell,m} a_{\ell {\rm m}} (\nu) \,
Y_{\ell}^{\rm m}(\hat{\bm{n}}) \,,
\label{eq:alm}
\end{equation}
and the  MAPS is defined as
\begin{equation}
C_{\ell}(\nu_a, \nu_b) = \big\langle a_{\ell {\rm m}} (\nu_a)\, a^*_{\ell
  {\rm m}} (\nu_b) \big\rangle\,
\label{eq:cl}
\end{equation}
where $\langle  ... \rangle$ denotes an ensemble average  over different statistically independent realizations of the random  field  $\delta T_{\rm b} (\hat{\bm{n}},\,\nu)$. 

The details of the visibility based TGE for measuring the MAPS and the PS are given in \citetalias{Bh18} and \citetalias{Pal20}. Here, we briefly summarize the mathematical formalism for this estimator. Starting from the visibility data, $\mathcal{V}_{i}(\nu_{a})$ corresponding to the i-th baseline $\textbf{U}_{i}$ and frequency $\nu_{a}$, the TGE first convolves the measured $\mathcal{V}_{i}(\nu_{a})$ with  $\tilde{w}(\u)$ which is the Fourier transform of a  window function ${\cal W}(\theta)$ suitably chosen to taper the primary beam (PB) of the telescope far away from the phase center. We divide the uv plane in a rectangular grid and the convolved visibilities $\V_{cg}$ at the grid-point $\u_{g}$ is given by,
\begin{equation}
\V_{cg}(\nu_{a}) = \sum_{i}\tilde{w}(\u_g-\u_i) \, \V_i(\nu_{a}) \,F_i(\nu_a).
\label{eq:a1}
\end{equation}
Here the subscript `$a$' denotes an individual channel  and $a=1,2....N_{c}$ where $N_{c}$ is the total number of channels that cover a bandwidth $B_{bw}$. The factor $F_i(\nu_a)$ is used to incorporate the flagging information. $F_i(\nu_a)$ is assigned a value `$0$' if the data at the baseline $\textbf{U}_{i}$ and frequency $\nu_{a}$ is flagged and $F_i(\nu_a)$ is `$1$' otherwise.

We use a Gaussian window function, ${\cal W}(\theta)=e^{-\theta^{2}/\theta^{2}_{w}}$, to taper the sky signal away from the phase center.  The main lobe of the PB of any telescope with a circular aperture can be approximated as, $\mathcal{A}(\theta)=e^{-\theta^{2}/\theta^{2}_{0}}$ where $\theta_{0}\sim 0.6 \times \theta_{\rm FWHM}$, $\theta_{\rm FWHM}$ being the full width at half maxima of $\mathcal{A}(\theta)$\citep{BS01, samir14}. We choose $\theta_{w}=f\theta_{0}$, where `$f$' represents the tapering parameter and controls the degree to which the PB pattern is tapered. The convolution implemented in eq.~(\ref{eq:a1}) equivalently amounts to modulating $\mathcal{A}(\theta)$ with ${\cal W}(\theta)$ by a multiplication, where $f>1$ will provide very little tapering and $f<1$ can be used to highly suppress the PB away from the phase center.

We define the TGE for MAPS in \citetalias{Bh18} as,
\begin{align}
\hat{E}_g(\nu_a,\nu_b) &= M_g^{-1}(\nu_a,\nu_b) {\mathcal Re} \Big[\V_{cg}(\nu_a)   \V_{cg}^{*}(\nu_b) \nonumber \\ 
& - \delta_{a,b} \, \sum_i F_i(\nu_a)  \mid
\tilde{w}(\u_g-\u_i) \mid^2   | \V_i(\nu_a) |^2  \Big] 
\label{eq:TGEI}
\end{align}

where ${\mathcal Re}[..]$ refers to real part of the expression within the brackets [..] and $M_g(\nu_a,\nu_b)$ is a normalization constant (discussed in detail later in this section). Along with the sky signal, each visibility $\mathcal{V}_{i}(\nu_{a})$ contains an additive   noise component $\mathcal{N}_{i}(\nu_{a})$ that is assumed to be a Gaussian random variable with zero mean and variance $ 2 \sigma_N^2$. The  noise in different baselines, frequency channels and timestamps are uncorrelated. Consequently, the noise contribution in MAPS is restricted  only to the self-correlations of the visibilities
\begin{equation}
    \langle \mathcal{N}_{i}(\nu_{a}) \mathcal{N}_{j}^{*}(\nu_{b}) \rangle =  \delta_{i,j}\delta_{a,b} 2 \, \sigma_N^2 \,.
\end{equation}
The   second term in the square brackets $[...]$  in eq.~(\ref{eq:TGEI}) subtracts out the contribution from the   self correlation of a visibility {\it i.e.} same baseline, frequency channel and timestamp. This exactly cancels out the noise contribution in the first term, and we obtain    an unbiased estimate of  MAPS.\\

We have validated eq.~(\ref{eq:TGEI}) (hereafter referred to as TGE-I) in \citetalias{Bh18} using realistic $150\, {\rm MHz}$ GMRT simulations. We have shown there that in the absence of foregrounds TGE-I can recover an input model 21-cm PS  with a very high accuracy even in the presence of noise and $80\%$ flagging in the visibility data.  However, the data available from the past and current $21$-cm experiments are dominated by various foregrounds  that overshadow the noise and the $21$-cm signal  by a few orders of magnitude \citep{ghosh1,ghosh2}. We find (shown later) that $C_{\ell}(\nu_a,\nu_b)$ estimated by applying TGE-I to such foreground dominated data  shows a discontinuity at $\nu_{a}=\nu_{b}$.
This discontinuity arises due to the self-correlation term  which is subtracted out only for $\nu_a=\nu_b$ in TGE-I.  We also find that this discontinuity  introduces a negative bias in the estimated PS $P(k)$. To deal with this problem, we have slightly modified TGE-I  in \citetalias{Pal20} to  obtain  
\begin{align}
\hat{E}_g(\nu_a,&\nu_b) = M_g^{-1}(\nu_a,\nu_b) {\mathcal Re} \Big[\V_{cg}(\nu_a)   \V_{cg}^{*}(\nu_b) \nonumber \\ 
& - \sum_i F_i(\nu_a)F_i(\nu_b)  \mid \tilde{w}(\u_g-\u_i) \mid^2   \V_i(\nu_a) \V_i^{*}(\nu_b)  \Big]  
\label{eq:a4}
\end{align}
where all the terms  hold the same meaning as in TGE-I. The modified estimator in eq.~(\ref{eq:a4}) (hereafter referred to as TGE-II) differs from TGE-I in the second term within the square brackets. This term now subtracts out the self-correlation of a visibility 
with itself {\it i.e.} same baseline and timestamp considering all possible combinations of frequencies $\nu_a$ and $\nu_b$. 
 This removes the discontinuity at $\nu_{a}=\nu_{b}$ in the estimated MAPS, and also avoids the negative bias in the estimated $P(k)$. We shall demonstrate this later in Section \ref{s4}, and we have validated TGE-II using simulations in Section~\ref{sec:simulation}.

We now consider the normalization factor $M_g(\nu_a,\nu_b)$. Here 
we have used  simulations to estimate the value of  $M_g(\nu_a,\nu_b)$. We first simulate multiple realizations of a Gaussian random field having unit multi-frequency angular power spectrum (UMAPS; $C_{\ell}(\nu_a, \nu_b)=1$). We use this as the sky signal to simulate the corresponding visibilities $ [\V_i(\nu_a)]_{\rm UMAPS}$ at the baselines and frequency channels identical to the data. The flagging of the actual data $F_i(\nu_a)$ was applied to the simulated visibilities  $ [\V_i(\nu_a)]_{\rm UMAPS}$ and these are then analyzed identically to the actual data to obtain 

\begin{align}
{M}_g & (\nu_a,\nu_b) = {\mathcal Re} \Big[\V_{cg}(\nu_a)   \V_{cg}^{*}(\nu_b) \nonumber \\ 
& - \sum_i F_i(\nu_a)F_i(\nu_b)  \mid \tilde{w}(\u_g-\u_i) \mid^2   \V_i(\nu_a) \V_i^{*}(\nu_b)  \Big]_{\rm {UMAPS}}.  
\label{eq:a4a}
\end{align}
We average over multiple realizations of the simulated UMAPS to reduce the statistical uncertainties in the estimated values of $M_g(\nu_a,\nu_b)$. For the subsequent analysis, we have simulated $50$ realizations of UMAPS and used these to estimate $M_{g}$. Note that our estimator does not incorporate the migration of the baselines with frequency and considers the values of the baselines to be ﬁxed at the reference frequency $\nu_{c}$.

The estimator in eq.~(\ref{eq:a4}) gives an unbiased estimate of the MAPS  $\langle {\hat E}_g(\nu_a,\nu_b) \rangle =C_{\ell_g}(\nu_a,\nu_b)$  at the grid point $\u_g$, or equivalently at angular multipole  $\ell_g=2\,\pi\,\mid \u_g \mid$. 
To increase the signal-to-noise ratio, we  further bin the entire $\ell$ range into $10\,\,\ell$ bins. The bin averaged Tapered Gridded Estimator is defined as, 
\begin{equation}
{\hat E}_G[q](\nu_a,\nu_b) = \frac{\sum_g w_g  {\hat E}_g(\nu_a,\nu_b)}
{\sum_g w_g } \,,
\label{eq:a6}
\end{equation}
where the sum is over all the grid points $\u_g$ in the  $q$'th bin   and  the $w_g$'s are the corresponding weights. Here, we have used $w_g=M_g(\nu_a,\nu_b)$  which implies that the weight is proportional to the baseline density of the particular grid point. The ensemble average of ${\hat E}_G[q](\nu_a,\nu_b)$ gives an unbiased estimate of the bin averaged MAPS $\bar{C}_{\bar{\ell}_q} (\nu_a,\nu_b)$ at the effective angular multipole $\bar{\ell}_q = \frac{ \sum_g w_g \ell_g}{ \sum_g w_g}$. Throughout this work we have considered baselines within $U\le3000\lambda$ (equivalently, $\ell\le18850$) and divided this into $10\,\,\ell$ bins. The effective angular multipoles corresponding to these bins cover a range $535 \lesssim \bar{\ell}_q \lesssim 15850$. Note that $\bar{\ell}_q$ vary slightly with the value of  $f$  and the values quoted in this paper have been estimated at$f=0.6$.  In the subsequent discussion we have used the simplified notation $C_{\ell}(\nu_{a},\nu_{b})$ and $\ell$ to denote $\bar{C}_{\bar{\ell}_q}(\nu_a,\nu_b)$ and $\bar{\ell}_q$ respectively.

Considering a  sufficiently small bandwidth of observation,  the redshifted $21$-cm signal can be assumed to be statistically homogeneous (ergodic) along the line-of-sight   (e.g. \citealt{Mondal19}). This   allows us to express $C_{\ell}(\nu_a,\nu_b)$ in terms of $\cl$ where $\Delta \nu = \mid \nu_b-\nu_a \mid$. This means that the statistical properties of the signal can now be entirely described as a function of the frequency separations $\Delta \nu$. Under the flat sky approximation, $P(k_{\perp}, k_{\parallel})$ the 3D power spectrum of the 21-cm brightness temperature fluctuations 
is then given by the Fourier transform of $\cl$ along the line-of-sight \citep{KD07},
\begin{equation}
P(k_{\perp},\,k_{\parallel})= r^2\,r^{\prime} \int_{-\infty}^{\infty}  d (\Delta \nu) \,
  e^{-i  k_{\parallel} r^{\prime} \Delta  \nu}\, C_{\ell}(\Delta \nu)
\label{eq:cl_Pk}
\end{equation}
where $k_{\parallel}$ and $k_{\perp}=\ell/r$ are the components of $\kv$ respectively parallel and perpendicular to the line-of-sight, $r$ and $r^{\prime}=dr/d\nu$ are respectively the comoving distance and its derivative with respect to $\nu$, both evaluated at the reference frequency $\nu_{c}=432.8\,{\rm MHz}$.   Here $r$ and $r^{\prime}$ are evaluated to have values  $5703\,{\rm Mpc}$ and $9.85\,{\rm Mpc/MHz}$ respectively.

We use a maximum likelihood estimator to estimate the PS $\bar{P}(k_{\perp},k_{\parallel m})$ from the measured  $C_{\ell}(n \, \Delta\nu_{c})$, where $n,\,m\, \epsilon\, [0,N_{E}-1]$ and $N_{E} = N_{c}/2$. Note that here we have used half of the available frequency separations $0\le\Delta\nu\le (N_{c}/2-1)\Delta\nu_{c}$ to avoid the poorly sampled higher frequency separations. In matrix notation,
\begin{equation}
    C_{\ell}(n \, \Delta\nu_{c})=  \sum_{m} \textbf{A}_{nm} \, \bar{P}(k_{\perp},k_{\parallel m}) + [\textrm{Noise}]_{n}
\label{eq:b2}
\end{equation}
where $\textbf{A}_{nm}$ are the components of the $N_{E} \times N_{E}$  Hermitian matrix $\textbf{A}$ containing the coefficients of the Fourier transform and $[\textrm{Noise}]_{n}$ is an additive noise associated with each estimated $C_{\ell}(n \, \Delta\nu_{c})$.
The maximum likelihood estimate of $\bar{P}(k_{\perp},k_{\parallel m})$ is given by, 
\begin{align}
\bar{P}(k_{\perp},k_{\parallel m}) =  \sum_n \{ [\textbf{A} ^{\dagger} \textbf{N}^{-1} \textbf{A}]^{-1} \textbf{A}^{\dagger}   & \textbf{N}^{-1} \}_{mn} \nonumber\\ &\{\mathcal{W}_{\rm BN}(n\Delta\nu_{c}) C_{\ell}(n\Delta\nu_{c})\}
\label{eq:ML}
\end{align}
where $\textbf{N}$ is the noise covariance matrix and `$\dagger$' denotes  the Hermitian conjugate. We have also introduced a Blackman-Nuttall (BN; \citealt{nut81}) window function $\mathcal{W}_{\rm BN}(n\Delta\nu_{c})$ along the $\Delta\nu$ to reduce any unwanted ripples in the estimated PS along $k_{\parallel}$ arising due to the finite bandwidth of observation.

We have estimated $\textbf{N}$ through `noise-only' simulations. As mentioned earlier, we have assumed that the noise in the visibilities are drawn from a  Gaussian random distribution with zero mean and variance $2 \sigma^{2}_{N}$,   and are uncorrelated at different baselines, frequencies and timestamps. We have simulated visibilities corresponding to the system noise only at baselines and frequencies identical to the actual data along with the flagging statistics. We use the value of $\sigma^{2}_{N}$  estimated from the data itself (Table \ref{tab:flagandrms}). We apply our estimator (eq.~\ref{eq:a4})  to estimate the MAPS corresponding to the simulated noise only visibilities. We generate multiple statistically independent noise-only visibility realizations to estimate the noise covariance matrix $\textbf{N}$ from the estimated MAPS. This method has been validated in \citetalias{Pal20}. The reader is referred to \citetalias{Pal20} for further details. Throughout the work, we have used $50$ noise realizations to estimate the noise covariance matrix $\textbf{N}$. Further, we have also estimated the PS  $P(k_{\perp}, k_{\parallel})$ for each of these noise-only simulations, and we have determined the mean and variance $[\delta P_{N}]^{2}$  of these values. As expected, the mean is consistent with zero. We have used  $[\delta P_{N}]$ to quantify the system noise contribution to the statistical fluctuations of the estimated PS of the actual data.

We use eq.~(\ref{eq:ML}) to estimate of the 3D PS $\bar{P}(k_{\perp},k_{\parallel m})$ from the measured $C_{\ell}(n\Delta\nu_{c})$. We have further binned $\bar{P}(k_{\perp},k_{\parallel m})$ along $k_{\parallel m}$ to obtain the bin averaged  $P(k_{\perp}, k_{\parallel})$ which we present in the subsequent analysis. The estimated bin averaged cylindrical power spectra $P(k_{\perp}, k_{\parallel})$ span a $(k_{\perp}, k_{\parallel})$  range of $0.09\le k_{\perp} \le2.78\,\,{\rm Mpc}^{-1}$ and $0\le k_{\parallel} \le13.1$ Mpc$^{-1}$ respectively.

\subsection{A comparison between TGE-I and TGE-II}
\label{s4}

% MAPS for two estimators for two l-values
\begin{figure}
\begin{center}
\includegraphics[width=\columnwidth]{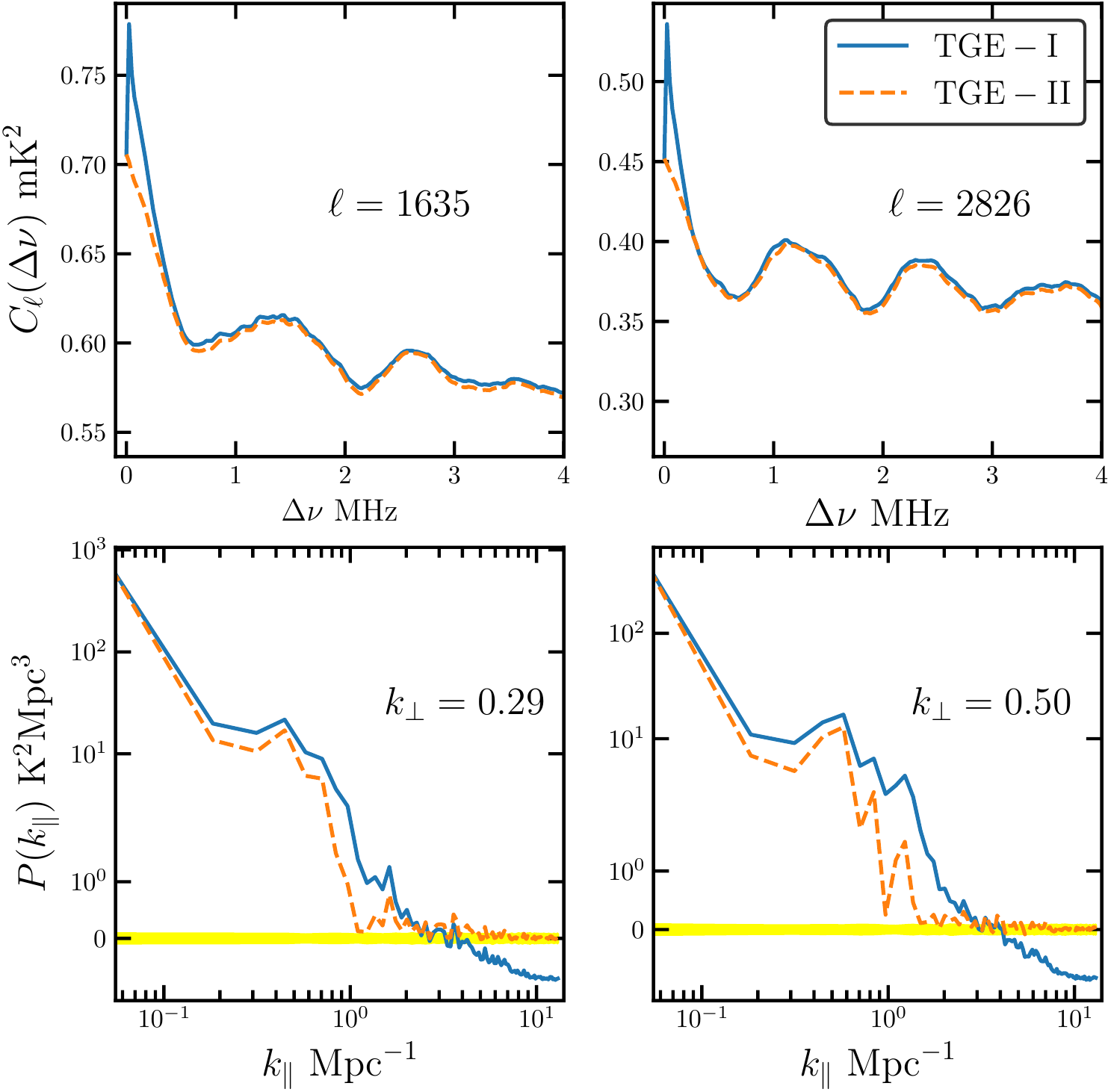}
\caption{A comparison of TGE-I (blue solid lines)  with TGE-II (orange dashed lines) applied on the combined nights data with $f=0.6$.  The upper panels show $C_{\ell}(\Delta\nu)$ as a function of $\Delta\nu$ for two different $\ell$ values.  The lower panels show slices of the estimated $P(k_{\perp}, k_{\parallel})$ as a function of $k_{\parallel}$ at a fixed value of $k_{\perp}$ corresponding to the $C_{\ell}(\Delta\nu)$ shown in the upper panels.  The yellow shaded regions show  $1\sigma$ errors $\delta P_{N}$ due to  system noise.}
\label{fig:comp}
\end{center}
\end{figure}

In this sub-section, we briefly demonstrate the shortcoming of TGE-I (eq.~\ref{eq:TGEI}) which was originally defined in \citetalias{Bh18}, and we also show that these can be overcome by TGE-II (eq.~\ref{eq:a4}) which we have used here. We apply both estimators to the combined nights data for the tapering parameter $f=0.6$. The upper panels of Figure~\ref{fig:comp} show the estimated $C_{\ell}(\Delta\nu)$ as a function of $\Delta\nu$ for two $\ell$ values  for both  TGE-I and TGE-II. We have restricted the frequency range to $4\,{\rm MHz}$ in the figure to highlight the abrupt discontinuity observed at $\Delta\nu=0$ for $C_{\ell}(\Delta\nu)$ estimated with TGE-I. As discussed in Section \ref{s3}, this dip arises due to the self-correlation term  which is only subtracted for $\nu_{a}=\nu_{b}$  to remove the noise bias. We see that the discontinuity is not present for TGE-II where we have subtracted out the self-correlation  at all $\Delta\nu$. We also note that, as expected,  the results from TGE-I and II both match for large $\Delta \nu$.

The lower panels of Figure~\ref{fig:comp} show slices of the PS  $P(k_{\perp}, k_{\parallel})$ along $k_{\parallel}$ at a fixed $k_{\perp}$ estimated using eq.~(\ref{eq:ML}) from  the $C_{\ell}(\Delta\nu)$ shown in the upper panels.  The yellow shaded regions show the $1\sigma$ statistical fluctuations ($\delta P_{N}$) arising due to system noise. In all cases $P(k_{\perp}, k_{\parallel})$ has relatively large values at small $k_{\parallel}$ which correspond to modes within the foreground wedge.  The values of $P(k_{\perp}, k_{\parallel})$ fall with increasing   $k_{\parallel}$ up to $k_{\parallel}\sim4\,{\rm Mpc}^{-1}$ beyond which the results from the two estimators are quite  different. For TGE-I (blue solid line), in both the  panels we notice  that $P(k_{\perp}, k_{\parallel})$ has negative values for $k_{\parallel} \gtrsim  4\,{\rm Mpc}^{-1}$ and the values fall to  $\sim -0.8\,{\rm K}^{2}{\rm Mpc}^{3}$ at the largest  $k_{\parallel}$-bins. In contrast, we find that this  negative bias is absent in TGE-II (orange dashed lines) where the values of $P(k_{\perp}, k_{\parallel})$ oscillate around zero for  $k_{\parallel}\gtrsim 4\,{\rm Mpc}^{-1}$. Further, we also see that these oscillations are roughly within the yellow shaded region, indicating that these are consistent with the fluctuations expected from the system noise in the data. The negative bias in TGE-I arises from the abrupt dip at $\Delta \nu=0$ seen in $C_{\ell}(\Delta\nu)$. 
We have also noticed large negative values in $P(k_{\parallel})$ at a few grid points $\textbf{U}_{g}$ for TGE-II near the wedge boundary. This is mostly originating due to a combination of corrupted baselines (most possibly due to bandpass calibration errors) at a few grid points. At these grid points we also observe a small dip near $\Delta \nu=0$ in $C_{\ell}(\Delta\nu)$. At this stage, we have decided to flag these grid points, favouring less data rather than bad data. Typically about $\sim 17 \%$ of grid points are flagged. We see that the negative bias is not present for TGE-II after the flagging (Figure~\ref{fig:comp}), and we use the rest of the grid points for further analysis. We also drop the suffix ``-II'' and refer to this as TGE throughout the rest of the Paper.

\section{Simulation}
\label{sec:simulation}
\begin{figure}
\begin{center}
\includegraphics[width=\columnwidth]{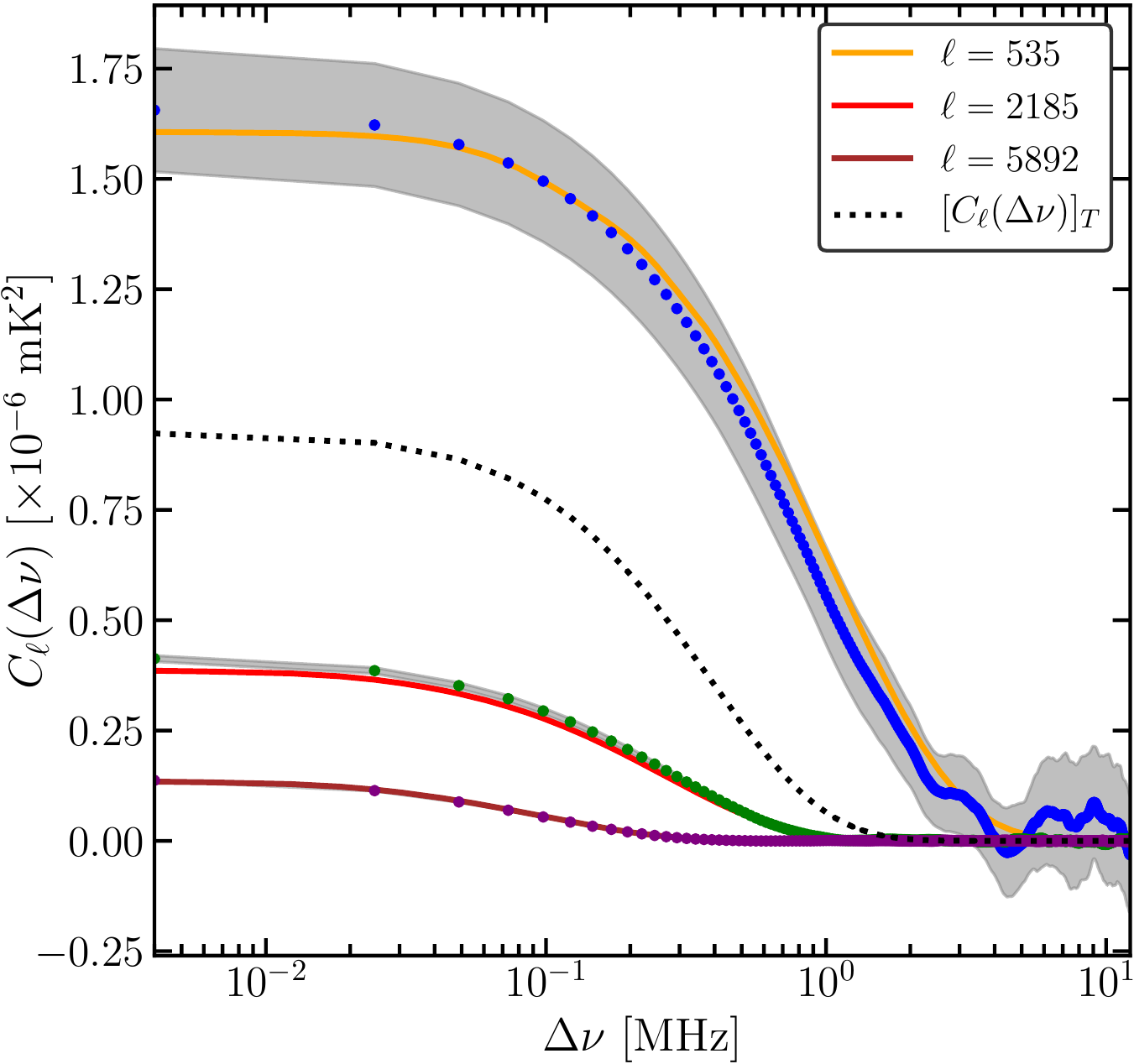}
\caption{The data points show the mean $C_{\ell}(\Delta\nu)$ with $2\,\sigma$ errors (shaded region) estimated from 16 realizations of the simulated sky signal. We have restricted $\Delta\nu$ to $\le 12.2\,\,{\rm MHz}$ in the plot, which we have used to estimate the PS. The solid lines show the analytical predictions corresponding to the input model  $P^m(k)$. The dotted line shows $[C_{\ell}(\Delta \nu)]_T$ the cosmological 21-cm signal predicted at $\ell=1635$ for the $\Lambda$CDM  model with $[\Omega_{\rm \HI} b_{\rm \HI}]=10^{-3}$.  Note that the $\Delta \nu=0$ points are shifted slightly for plotting on a logarithmic scale.}
\label{cl}
\end{center}
\end{figure}
\begin{figure}
\begin{center}
\includegraphics[width=\columnwidth]{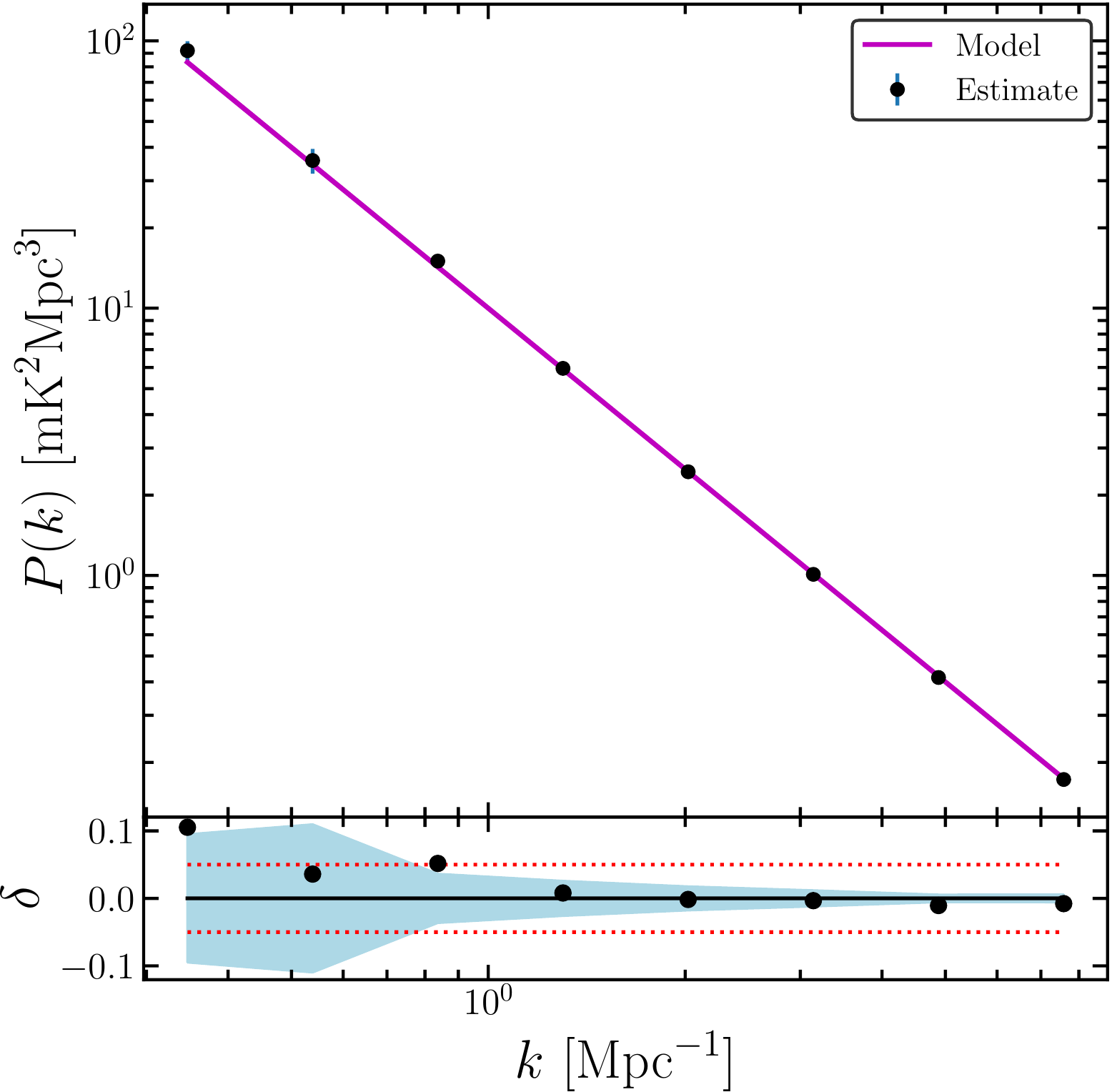}
\caption{The upper panel shows the estimated spherically-binned power spectrum $P(k)$ (data points) and $2\,\sigma$ error-bars for the simulations along with the input model $P^{m}(k)$ (purple solid line). The bottom panel shows the fractional deviation  $\delta=[P(k)-P^m(k)]/P^m(k)$ (data points) and the expected $2 \sigma$ statistical fluctuations for  the same (blue shaded region). The (red) dotted lines demarcate  the region where $\mid \delta \mid \le 0.05 $.}
\label{fig:pk_val}
\end{center}
\end{figure}

We have already validated TGE in \citetalias{Pal20} using $150\,{\rm MHz}$ GMRT simulations where $\sim47\%$ of the data were flagged. We found that TGE could recover the input PS with $<8\%$ fractional deviation over the entire $k$-range used for the analysis. In the present work, we have repeated a similar analysis for the sub-band 2 data, which we have analyzed here. The aim is to validate the estimator and quantify the accuracy to which TGE is expected to recover $P(k)$   for the data analyzed here.

The 21-cm brightness temperature fluctuations  $\delta T_{\rm b} (\hat{\bm{n}},\,\nu)$ in the simulations are assumed to be a Gaussian random field corresponding to an input  model  
\begin{equation}
    \label{eq:modelps}
    P^{m}(\bm{k})  = A \left( \frac{k}{k_0}\right)^n {\rm  mK^{2} \,  Mpc^{3}}\,.
\end{equation}
where we have arbitrarily set $A=10$, $k_0 = 1 \, \mathrm{Mpc}^{-1}$, and used a power law index $n=-2$. The simulations  closely follow the prescriptions of \cite{samir17} and \citetalias{Pal20}. We have carried out simulations on a $N^{3}=[1024]^3$ cubic grid with a grid spacing $\Delta L = 0.24$ Mpc which  matches  the spectral resolution $\Delta\nu_{c} = 24.4 \, {\rm kHz}$ of our data ($\Delta L = r^{\prime} \times \Delta\nu_{c}$). This   results in an angular  resolution  of $\Delta\theta \sim 8.4^{''}$ ($\Delta L = r \Delta\theta$), and  the angular  extent of the simulation box ($N\Delta\theta$) covers  $\sim2.5$ times the $\theta_{\rm FWHM}$ of GMRT at the frequency $\nu_c = 432.8 \, {\rm MHz}$. We have converted the simulated images into visibilities using the baseline distribution of the combined nights data. The simulations  incorporate the frequency dependence of the PB and  baseline migration.

We have applied the TGE (eq.~\ref{eq:a4}) on the simulated visibilities, and analyzed the simulated data identical to the actual data, to estimate the MAPS $C_{\ell}(\Delta\nu)$.  We have used $N_r=16$  independent realizations of the simulation to estimate the  mean $C_{\ell}(\Delta\nu)$ and the $2\,\sigma$ errors shown in Figure~\ref{cl} at three values of  $\ell$ for $f=0.6$. We have also shown (solid lines) the analytical model predictions  $C^m_{\ell}(\Delta\nu)$ calculated using  \citep{KD07,ali14} 
\begin{equation}
    C_{\ell}(\Delta\nu) = \frac{1}{\pi r^2} \int_{0}^{\infty}  d k_{\parallel} \cos(k_{\parallel}r^{\prime}\Delta\nu) P(\bm{k})\,.
    \label{eq:CT}
\end{equation}
We see that the $C_{\ell}(\Delta\nu)$  estimated from the simulations closely matches  the  analytical  prediction $C^m_{\ell}(\Delta\nu)$ which are mostly within the shaded region showing the $2\,\sigma$ uncertainty. The deviations  between $C_{\ell}(\Delta\nu)$ and $C^m_{\ell}(\Delta\nu)$ are found to lie within $\lsim\, 10\%$ at $\Delta \nu =0$. These deviations are primarily due to uncertainties in the normalization factors $M_{g}(\nu_{a},\nu_{b})$ which have been estimated from $50$ UMAPS realizations. To test this we have checked  that these deviations decrease  if  the number of UMAPS realizations is increased.  We also note that the large $\Delta\nu$ are poorly sampled compared  to the small $\Delta\nu$, and the cosmic variance increases as we go to larger frequency separations.

The dotted line in Figure~\ref{cl}  shows $[C_{\ell}(\Delta \nu)]_T$ an estimate of the cosmological 21-cm signal expected in the observed data. It is assumed that the fluctuations of  the \HI distribution trace the  underlying matter distribution with a 
linear bias $b_{\rm \HI}$. This allows us to express   $P_{T}(\bm{k})$    the predicted 21-cm brightness temperature power spectrum in terms of  $P^s_m(\bm{k})$  the underlying matter power spectrum in redshift space.  Here we have rewritten 
eq.~(23)  of \cite{BA5} as 
\begin{equation}
    P_{T}(\bm{k}) \, = \,[\Omega_{\rm \HI} b_{\rm \HI}]^{2}\, \bar{T}^{2}\, P^s_{m}(\bm{k})
    \label{eq:ph1}
\end{equation}
with 
\begin{equation}
    \bar{T}(z) = 133 \,{\rm mK}\,(1+z)^{2}\,\bigg(\frac{h}{0.7}\bigg)\,\bigg(\frac{H_{0}}{H(z)}\bigg)
    \label{eq:tbar}
\end{equation}
where the cosmological  \HI mass density $\Omega_{\rm \HI}$ is the comoving \HI mass density in units of the present critical density. 
DLA observations  (e.g. \citealt{Not, Zafar})  show that  $\Omega_{\rm \HI} \sim 10^{-3}$ across $1.5 < z < 5$,   whereas various simulations (e.g. \citealt{Deb16})  indicate $1 \le b_{\rm \HI} \le 2$ across $(2 \le z \le 3)$. For the estimates presented  here we have used $[\Omega_{\rm \HI} b_{\rm \HI}]=10^{-3} $  and a fitting formula for $P_m(k)$ \citep{Eisenstein_1998},  ignoring the effect of redshift space distortion. We find that $[C_{\ell}(\Delta \nu)]_T$ has a peak value of $\approx 0.9 \times 10^{-6} \, {\rm mK}^2$ at $\Delta \nu=0$ for the value of $\ell$ $(=1635)$  shown here. The value of $[C_{\ell}(\Delta \nu)]_T$ decreases with increasing $\Delta \nu$ and it is $\approx 0$ for $\Delta \nu > 1 \, {\rm MHz}$. In fact, a similar behaviour is also seen for the model predictions $[C_{\ell}^m(\Delta \nu)]$ where we find that the value  peaks at $\Delta \nu =0$ and decorrelates rapidly with increasing $\Delta \nu$ with a value $\approx 0$ at $\Delta \nu > 1 \, {\rm MHz}$. The peak value reduces with increasing $\ell$ for which the signal also decorrelates faster. These are generic features of the expected 21-cm signal \citep{Bharadwaj-Pandey-2003} irrespective of the details of the 21-cm PS.

We have implemented eq.~(\ref{eq:ML}) to estimate the PS of the simulated sky signal. Identical to the actual data, we have also used a BN window function along the frequency separation for these simulations. The simulations differ from the data in that the error-covariance $\textbf{N}$ is dominated by cosmic variance, whereas the data is system noise dominated. Here we have used the covariance of the simulated $C_{\ell}(\Delta\nu)$ to estimate the noise covariance matrix $\textbf{N}$. The upper panel of  Figure~\ref{fig:pk_val} shows the estimated spherically-binned PS $P(k)$ and the associated $2\,\sigma$ errors along with the model PS $P^{m}(k)$. We see that $P(k)$ is in reasonably good  agreement with $P^{m}(k)$ across the entire $k$ range considered here. The lower panel of Figure~\ref{fig:pk_val} shows the fractional deviation $\delta=[P(k)-P^{m}(k)]/P^{m}(k)$ and the expected $2\,\sigma$ statistical fluctuations for the same. We have $\mid \delta \mid \lsim\,\, 5 \%$ in most of the $k$-bins shown here. We have somewhat larger deviation $(\mid \delta \mid  \sim 10\%)$ at the smallest $k$-bin. The convolution with the window function (eq.~\ref{eq:a1}) is expected to become important at the small baselines \citep{samir14}, and this possibly contributes to enhance the deviations in the small $k$-bins. A part of the deviations could also arise from the low baseline density in some of the bins (Figure~\ref{fig:uv}). We see that the $\delta$ values are all consistent with the predicted $2 \sigma$ errors. In the analysis of the actual observed data, as presented later in this paper, we have identified some  of the $(k_{\perp},k_{\parallel})$ modes as being foreground contaminated. These modes have been  excluded for estimating the spherically-binned PS $P(k)$  of the actual data. In keeping with this, we have also excluded these modes for the simulations presented here. The entire validation presented here used {\it exactly}  the same $(k_{\perp},k_{\parallel})$ modes as those that have been used for the actual data.  In summary, we have validated the TGE and we find that it is able recover the input model PS to an accuracy better than $\lsim \,\, 10 \%$ across the entire $k$ range considered here, and $\lsim\,\,5\%$ across $0.54 \le k \le 7.58\,\,{\rm Mpc}^{-1}$. The results are not very different even if we include all the available $(k_{\perp},k_{\parallel})$ modes  to estimate $P(k)$.

\section{Results}
\label{s6}

\subsection{The Estimated MAPS}
\label{maps}

We have used the TGE (eqs.~\ref{eq:a4} and \ref{eq:a6}) to estimate the MAPS $C_{\ell}(\Delta\nu)$ from the calibrated and compact source subtracted visibility data for the individual nights of observation as well as the combined data. We see that the May 6 data (Table ~\ref{tab:flagandrms}) has the least flagging as well as the smallest visibility r.m.s. Guided by this,  we first consider the results for the May 6 data and subsequently use this as a reference for comparing the results for the other nights (not shown here)  and the combined data. The two polarizations (LL and RR) were treated as independent measurements from the same baseline. We have repeated the analysis for three values of the tapering parameter $f=5.0,\,2.0,\,\textrm{and}\,0.6$. As mentioned earlier, the tapering increases with decreasing value of $f$, and $f=5.0$ can be considered equivalent to an untapered PB.

\begin{figure}
\begin{center}
\includegraphics[width=\columnwidth]{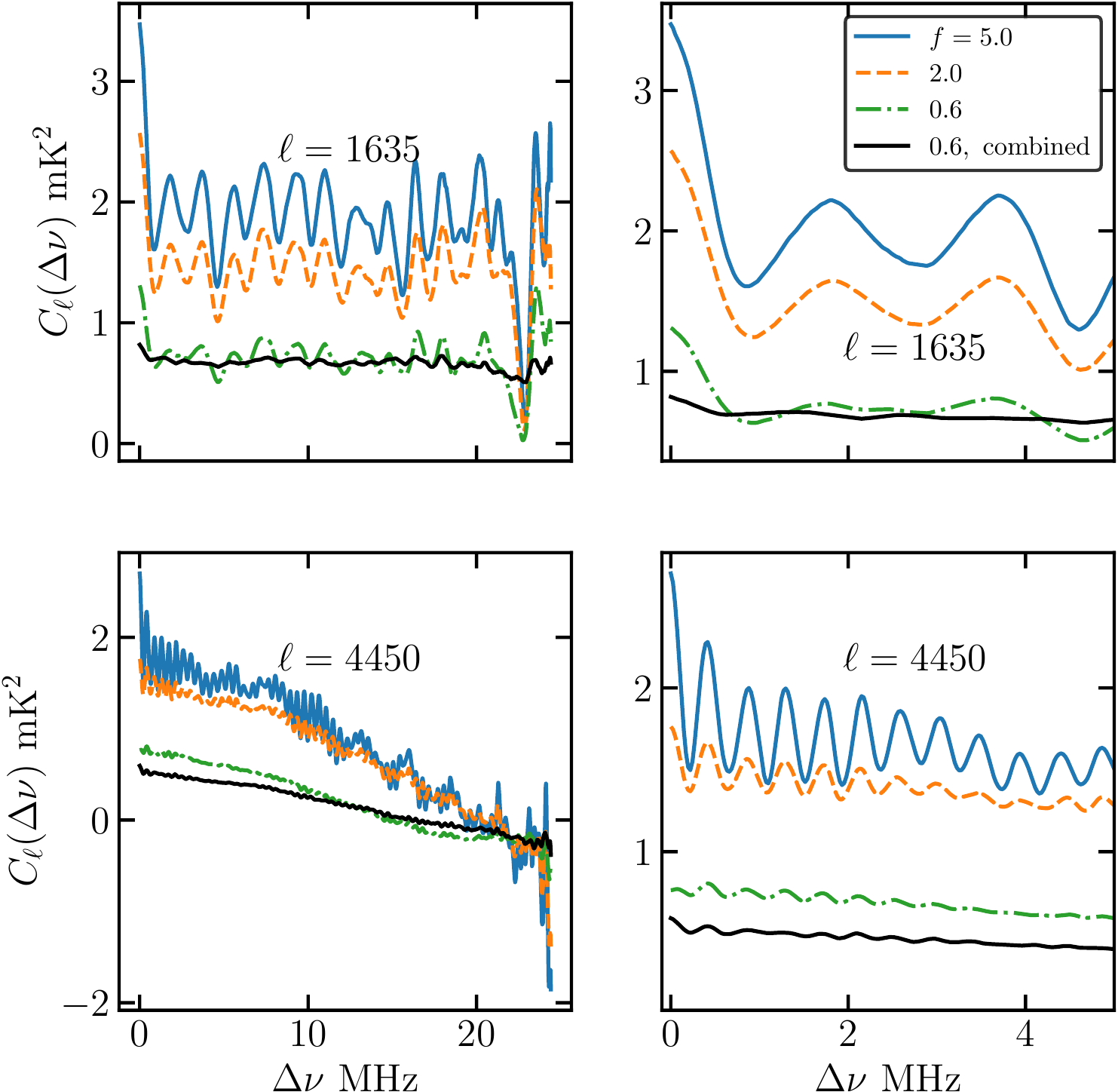} %bb=0 0 30 30
\caption{$C_{\ell}(\Delta\nu)$ as a function of $\Delta\nu$ for the May 6 observation are shown at $f=5.0,\, 2.0,\,{\rm and} \,0.6$ at two values of the angular multipole $\ell$. The left panels show the $C_{\ell}(\Delta\nu)$ for the entire $24.4\, {\rm MHz}$ bandwidth considered here. The right panels show the same but we restrict the frequency separation upto $5\, {\rm MHz}$. The black solid lines represent the estimated $C_{\ell}(\Delta\nu)$ using $f=0.6$ for the combined data.}
\label{fig:6may_f}
\end{center}
\end{figure}

In  Figure~\ref{fig:6may_f} we have shown $C_{\ell}(\Delta\nu)$ as a function of the frequency separation $\Delta\nu$ at different values of $f$ for the May 6 data. The upper and lower panels correspond to two representative $\ell$-values, $\ell=1635$ and $4450$ respectively. The left and right panels show $C_{\ell}(\Delta\nu)$ over the entire $24.4\,{\rm MHz}$ bandwidth and $\Delta\nu\le 5\,{\rm MHz}$ respectively. We see that  $C_{\ell}(\Delta\nu)$ exhibit an oscillatory pattern  whose frequency  increases with $\ell$. This increase in the frequency of oscillation is more evident in the right panels  which show  a small part  of the $\Delta\nu$ range.  These oscillatory patterns are consistent with the expected foreground behaviour \citep{ghosh1,ghosh2,ghosh3}. The contribution to $C_{\ell}(\Delta\nu)$ from  a single point source  is predicted  (\citetalias{Pal20})  to be 
\begin{equation}
    C_\ell (\Delta\nu) \propto \cos{(\ell \theta \Delta \nu/\nu_c)} \,
    \label{eq:oscillation}
\end{equation}
where $\theta $ is the sine of the angle between the source position and the phase center of the observation. As mentioned earlier, the compact and discrete sources within the main lobe of the PB have been modelled and subtracted out \citep{Cha2}. However, far-field residual sources remain that are difficult to model and clean out from the data. The oscillations in the estimated $C_{\ell}(\Delta\nu)$ is essentially a superposition of the oscillatory contributions from all the residual sources outside the main lobe of the PB. Note that the oscillation in $C_{\ell}(\Delta\nu)$ is fundamentally due to the chromatic nature of radio-interferometric measurements. The increase in the frequency of oscillation in $C_{\ell}(\Delta\nu)$ at larger baselines yields the `wedge' shape in the PS, which we shall present shortly. The extent of this `foreground wedge' is determined by the position ($\theta$) of the wide-field source,  which can maximally reach the horizon limit $\theta \sim 1$. The oscillations seen here, or equivalently the foreground wedge, arises due to `baseline migration' 
\citep{adatta10, Morales_2012, parsons12, vedantham12, Murray_2018}.  TGE allows us to suppress the antenna response at large angular distances relative to the phase center, reducing the large angular-scale foreground contributions present in the data. This is illustrated in Figure~\ref{fig:6may_f} where we see that for both the $\ell$ values the overall amplitude of $C_{\ell}(\Delta\nu)$ goes down as the value of $f$ is reduced (or equivalently, the tapering is increased). Comparing with respect to $f=5.0$, we find that   the amplitude of $C_{\ell}(0)$ drops by a factor $3-4$ for  $f=0.6$ at the $\ell$ values shown here. Further, the amplitude of the oscillatory pattern also decreases considerably as the value of $f$ is reduced. The implication of this on the PS  will be discussed in Section \ref{power} where we consider the PS for different values of $f$. 

\begin{figure}
\begin{center}
\includegraphics[width=\columnwidth]{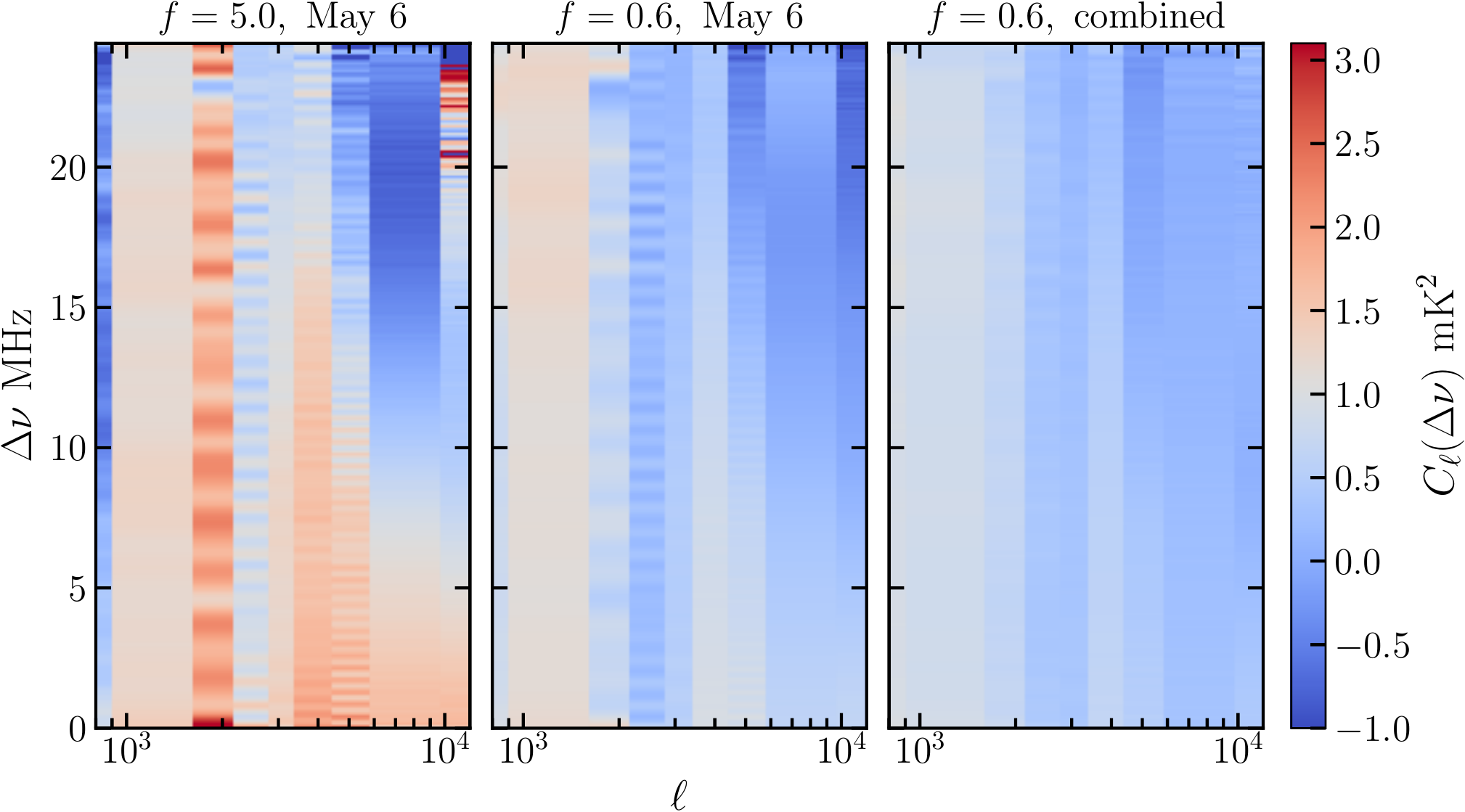}
\caption{This shows $C_{\ell}(\Delta\nu)$ across the entire ($\ell, \Delta\nu$) range for May 6 at $f=5$ (left panel), May 6 at $f=0.6$ (middle panel) and four nights combined data at $f=0.6$ (right panel).}
\label{fig:MAPS3D}
\end{center}
\end{figure}

Figure~\ref{fig:MAPS3D} shows  $C_{\ell}(\Delta\nu)$ across the entire $(\ell,\Delta\nu)$ range. A comparison of the results obtained from the May 6 data with  $f=5.0$ (left panel) and $f=0.6$ (middle panel) demonstrates the effect of tapering. Considering the left panel,  we see that the oscillations along $\Delta\nu$ are prominently visible in most of the $\ell$-bins. For larger $\ell$,  the oscillations become so rapid that we cannot discern them in the figure. The overall amplitude and that of the oscillatory pattern are both visibly reduced in the middle panel. The rightmost panel shows  $C_{\ell}(\Delta\nu)$ for the combined nights data for $f=0.6$. It is evident from the last two panels of the Figure~\ref{fig:MAPS3D} that combining different nights data yields a further smoothing of the oscillatory patterns. This is also illustrated in Figure~\ref{fig:6may_f} which shows  $C_{\ell}(\Delta\nu)$ for the combined data in black-solid lines for $f=0.6$.  We see, in both the Figures \ref{fig:6may_f} and \ref{fig:MAPS3D}, that the overall amplitude of $C_{\ell}(\Delta\nu)$  as well as the amplitude of the oscillations are smaller for the combined nights in comparison to the May 6 data. The reason is that the convolution (eq.~\ref{eq:a1}), which incorporates the tapering in TGE,  is sensitive to the baseline distribution \citep{samir14}. The baseline densities for May 6 and the combined nights are shown in the lower panels of Figure~\ref{fig:uv}. We see that the baseline density increases by a factor $\sim 3.5$ for the combined nights data. The uv-coverage is also considerably less patchy in comparison to the May 6 data. We expect the tapering to be more effective for the denser and more uniform baseline coverage of the combined nights. We see that this expectation is borne out in the estimated  $C_{\ell}(\Delta\nu)$. 

In addition to the rapid oscillations in $\cl$ (Figures~\ref{fig:6may_f} and \ref{fig:MAPS3D}) which arise from the residual compact source contribution due to baseline migration, $\cl$ also exhibits a gradual de-correlation {\it i.e.} the value of $\cl$ decreases as $\Delta \nu$ increases. We expect the intrinsic frequency spectrum of the compact sources to cause a smooth de-correlation of $\cl$. However, it is interesting to note that the values of $\cl$ do not fall monotonically with increasing $\Delta\nu$. We can observe this in the lower left panel of Figure~\ref{fig:6may_f} where the  amplitude of the oscillations in $\cl$ decreases till $\Delta\nu \sim 5 \, \rm{MHz}$, then  increases again up to $\Delta\nu \sim 10 \, \rm{MHz}$, and then decreases again. This modulation, we believe, arises because of the PB pattern which changes with frequency across the frequency bandwidth considered here. The position of the null points of the PB changes considerably with frequency, and this possibly causes the slow modulation seen in $\cl$.

\subsection{The Estimated PS}
\label{power}
%Cylindrical PS heat map for combined night data at all tapering parameters
\begin{figure*}
\begin{center}
\includegraphics[width=\textwidth]{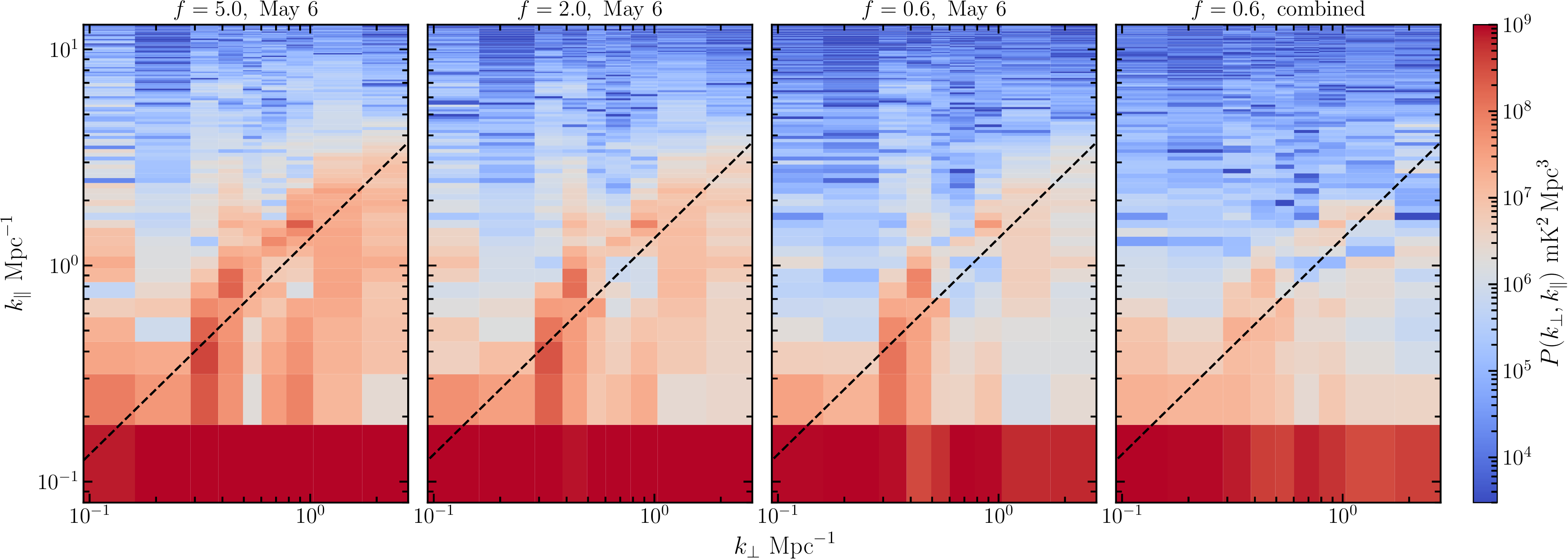}
\caption{The first three panels from the left show the absolute values of the cylindrical power spectra $P(k_{\perp},k_{\parallel})$ for the May 6 data for  different values of tapering. The rightmost panel shows the same for the combined nights data for  $f=0.6$. In all the cases the black dashed lines denote $[k_{\parallel}]_{H}$.}
\label{fig:Pk_f}
\end{center}
\end{figure*}

We have applied  the maximum likelihood method described in Section \ref{s3} on $C_{\ell}(\Delta\nu)$ to estimate the cylindrical power spectra $P(k_{\perp},k_{\parallel})$. 
The different panels of Figure~\ref{fig:Pk_f} show the absolute values of the PS.
The first three panels, starting from the left, respectively correspond to 
$f=5.0, \, 2.0, \, \rm{and} \, 0.6$  for the May 6 data while the right panel corresponds to $f=0.6$ for the combined nights.  In all cases,   the foregrounds are found to be largely confined within a wedge in the $(k_{\perp},k_{\parallel})$ plane. The wedge boundary (also called the `horizon limit') can be mapped to a straight line $[k_{\parallel}]_{H} = (r/r^{\prime}\nu_{c}) k_{\perp}$ in  the  $(k_{\perp},k_{\parallel})$ plane \citep{pober16}. The  region $k_{\parallel} \le [k_{\parallel}]_{H}$ is referred to as the ``foreground wedge'', and the PS estimated in this region of the  $(k_{\perp},k_{\parallel})$ plane is largely dominated by the foregrounds.  The region $k_{\parallel} > [k_{\parallel}]_{H}$ is relatively foreground-free, and we refer to this as the  ``21-cm  window''. While the bulk of the foregrounds in Figure~\ref{fig:Pk_f}  are localized within the foreground wedge, all the panels also show some foreground leakage outside the predicted wedge boundary. Various factors like the chromaticity of the sky signal and the PB, sparse sampling of baselines, calibration errors, etc., cause this foreground to leak into the 21-cm window \citep{bowman09, Thyagarajan_2016},  and it is often necessary to discard a part of the 21-cm window for estimating the PS of the 21-cm signal. 

We now compare the PS values in the three left panels of Figure~\ref{fig:Pk_f} which respectively correspond to $f=5.0, 2.0$ and $0.6$ for the May~6 data. We see that the overall foreground contamination comes down as the value of $f$ is reduced {\it i.e.} the tapering is increased, and we have a larger side-lobe suppression. In addition to a decrease in amplitude within the foreground wedge, we also notice a reduction in the leakage outside the wedge. However, this effect is not uniform across the different $k_{\perp}$-bins. The convolution which incorporates the tapering is expected to be more effective when we have a denser and more uniform $uv$ coverage of the baselines. As discussed in  Section $\ref{maps}$,  we have a denser baseline distribution at the small baselines, which is reflected in the fact that the foreground suppression is more effective at the lower $k_{\perp}$-bins. The rightmost panel of Figure~\ref{fig:Pk_f} corresponds to $f=0.6$ 
for the combined nights data. Comparing the two rightmost panels, both of which correspond to $f=0.6$, we see that we have a smaller foreground contribution for the combined nights data compared to the May 6 data. This is expected because the combined nights data has a $\sim 3.5$ times larger baseline density in comparison to the May 6 data (Figure \ref{fig:uv}). We see that we have the lowest level of foreground contribution for the combined nights data with $f=0.6$, and we have focused on this for the subsequent analysis of this paper.

% Section plots for combined night data at all tapering parameters
\begin{figure}
\begin{center}
\includegraphics[width=\columnwidth]{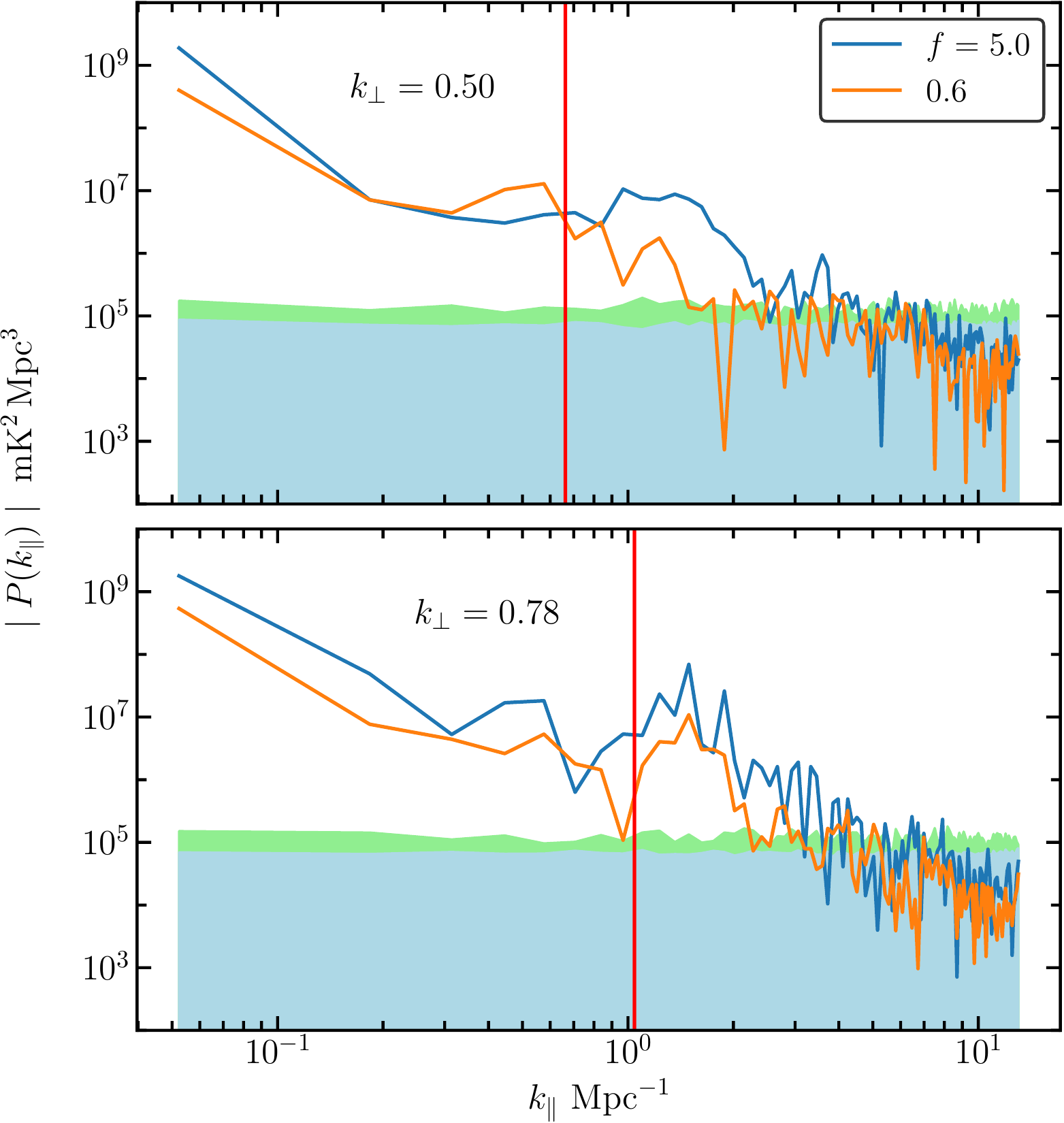}
\caption{The absolute values of the cylindrical power spectra $P(k_{\perp},k_{\parallel})$ for the combined nights data are shown as a function of $k_{\parallel}$ for  $f=5.0\,{\rm and}\,0.6$  at two representative values of $k_{\perp}$. The value at $k_{\parallel}=0$ has been slightly shifted for plotting on a log scale. The vertical red solid line denotes $[k_{\parallel}]_{H}$ at the respective $k_{\perp}$-bin. The green and light-blue shaded regions show the $1\sigma$ errors $[\delta P_{N}]$ due to the system noise estimated at $f=5.0\,{\rm and}\,0.6$ respectively.}
\label{fig:Pk_f_slice}
\end{center}
\end{figure}

We next consider  the estimated values of  $P(k_{\perp},k_{\parallel})$  in some detail. To study this,  in Figure~\ref{fig:Pk_f_slice} we have shown the absolute values  of $P(k_{\perp},k_{\parallel})$ as a function of $k_{\parallel}$ for two different fixed $k_{\perp}$ bins corresponding to  $k_{\perp}=0.50\,{\rm Mpc}^{-1}\,$ (upper panel) and $ \,0.78\,{\rm Mpc}^{-1}$ (lower panel). The results are shown for the combined nights data, considering two values of tapering namely $f=5.0$ and $f=0.6$. In each panel, the vertical red solid line  denotes $[k_{\parallel}]_{H}$ which is the predicted foreground wedge boundary for the particular $k_{\perp}$-bin. Further, the shaded region denotes the  predicted $1\sigma$ errors $([\delta P_{N}])$ due to the system noise contribution considering  $f=5.0$ (green) and $0.6$  (light-blue). In all cases we have the largest value of $P(k_{\perp},k_{\parallel})$  $(\sim 10^9 \, {\rm mK}^2 \, {\rm Mpc}^3)$ at $k_{\parallel}=0$. The value of $P(k_{\perp},k_{\parallel})$ falls $(\sim 10^6 \, - \, 10^7  \, {\rm mK}^2 \, {\rm Mpc}^3)$ till $k_{\parallel}$ approaches $[k_{\parallel}]_{H}$ where it flattens out,   and then  rises slightly in a few $k_{\parallel}$ bins  just beyond  $[k_{\parallel}]_{H}$. Further beyond this, the value of $\mid P(k_{\perp},k_{\parallel}) \mid $ falls  with increasing $k_{\parallel}$. We find  $\mid P(k_{\perp},k_{\parallel}) \mid  \sim 10^4 \, - \, 10^5  \, {\rm mK}^2 \, {\rm Mpc}^3$  at the largest   $k_{\parallel}$   bins where  the power oscillates between positive and negative values  which are comparable with the $1-\sigma$ error-bars.  We interpret the estimated power in this region to be arising due to a combination of system noise and some  residual foreground leakage. In both the panels we find that the values of  $\mid P(k_{\perp},k_{\parallel}) \mid $ decrease when $f$ is  reduced from $f=5.0$ to $f=0.6$.

Considering  Figure~\ref{fig:Pk_f_slice}, as noted earlier, we find  that   there is a $k_{\parallel}$ range within the foreground wedge  where $P(k_{\perp},k_{\parallel})$ has a relatively small value in comparison to the values at  $k_{\parallel}=0$ and $k_{\parallel} \approx [k_{\parallel}]_{H}$.   This is also seen in the various panels of  Figure~\ref{fig:Pk_f} where we see that the values of $P(k_{\perp},k_{\parallel})$  fall as we move away from $k_{\parallel}=0$ and then increases again at   $k_{\parallel} \approx [k_{\parallel}]_{H}$. We identify the rise in $P(k_{\perp},k_{\parallel})$ close to  the horizon limit as a wide-field foreground effect  known as the ``{\it pitchfork effect}'' \citep{thyag15_1,thyag15}.  
The pitchfork effect has been previously reported in observations with the MWA \citep{thyag15}, PAPER \citep{kohn16} and LOFAR \citep{gehlot17} telescopes at lower frequencies $(\sim 150 \, {\rm MHz})$ which target the redshifted 21-cm signal from the  Epoch of Reionization ($z>6$). The present work is possibly the first time this effect is being observed at higher frequencies which correspond to the post-reionization 21-cm signal. 
We notice (Figure~\ref{fig:Pk_f}) that the magnitude of the pitchfork effect is reduced as the tapering parameter is reduced from  $f=5$  to $f=0.6$. 

% Statistical analysis and limits from spherical binning

% Cylindrical PS heat map for combined nights at f=0.6
\begin{figure}
\begin{center}
\includegraphics[width=\columnwidth]{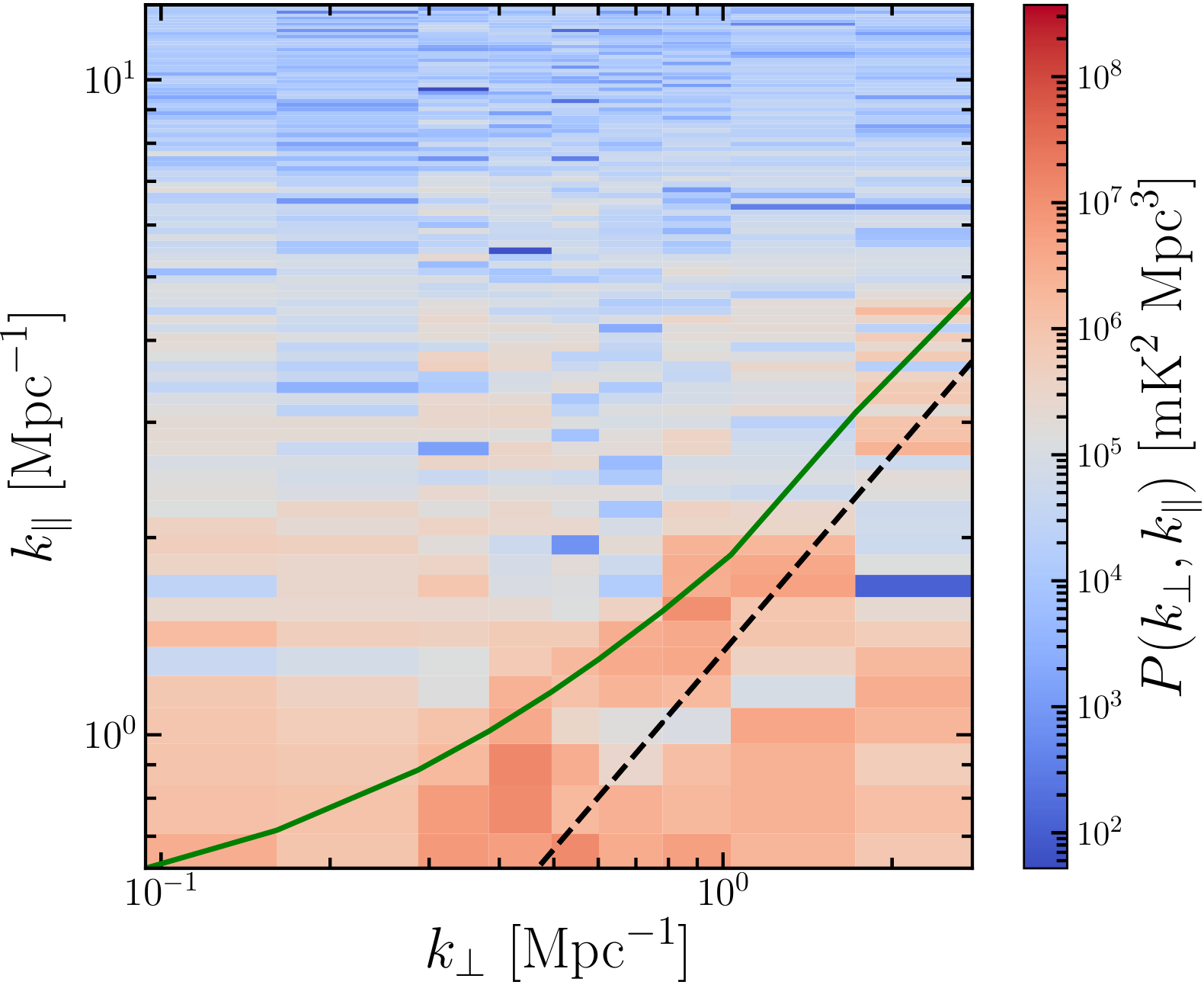}
\caption{The cylindrical power spectra $\mid P(k_{\perp},k_{\parallel}) \mid $ for the combined nights data for $f=0.6$. Here the black-dashed line denotes $[k_{\parallel}]_{H}$. The region above the green solid line has been used for spherical binning.}
\label{fig:pk-sph}
\end{center}
\end{figure}

\begin{figure}
\begin{center}
\includegraphics[width=85mm]{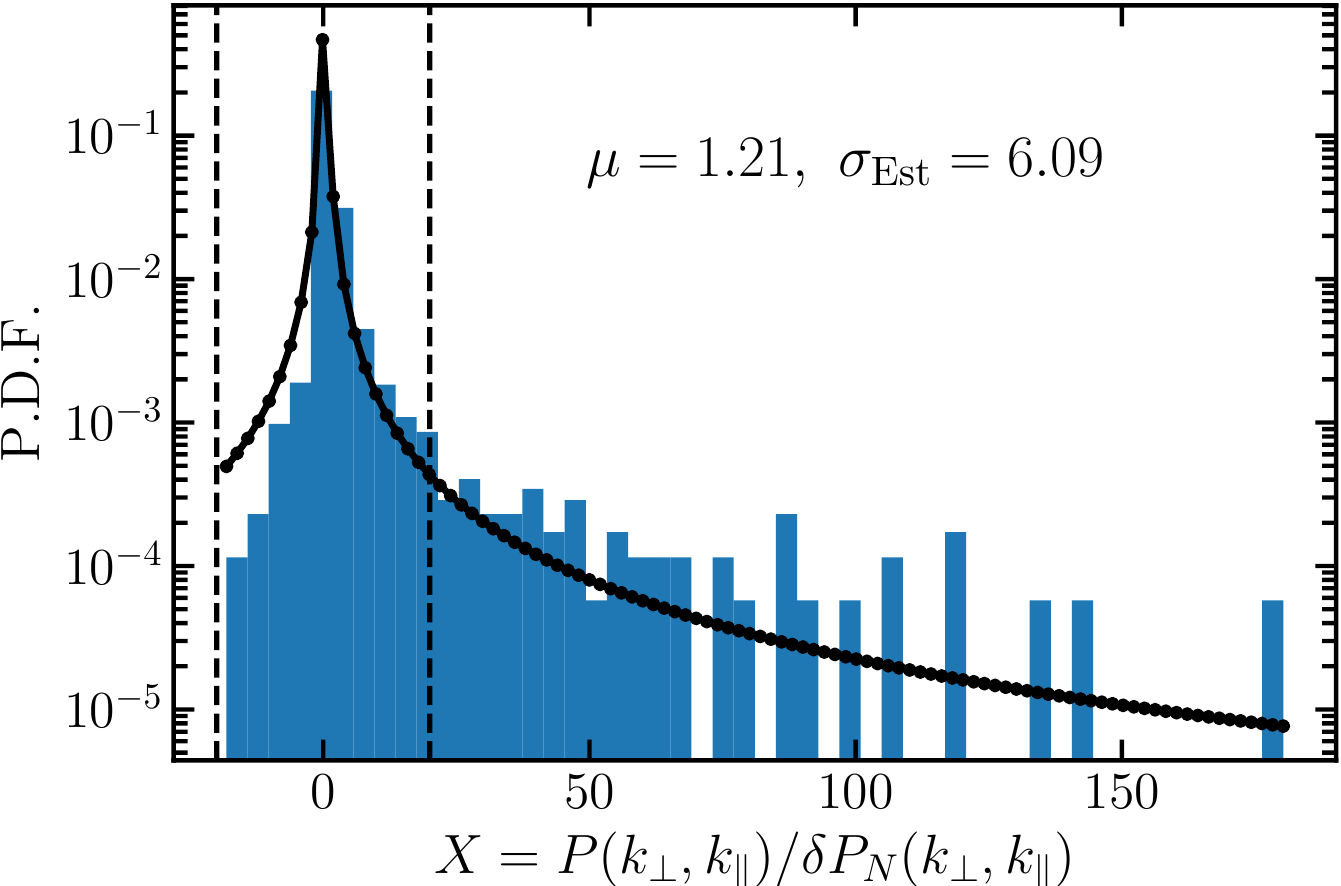}
\caption{The histogram of the variable $X=\frac{P(k_{\perp},\,k_{\parallel})}{\delta P_{N}(k_{\perp},\,k_{\parallel})}$. The black-solid curve shows the fit with t-distribution. The mean $1.21$ and standard deviation $6.09$ are obtained from $\mid X \mid \le 20$, demarcated by the vertical black-dashed lines.}
\label{pkbysigma}
\end{center}
\end{figure}

Figure~\ref{fig:pk-sph} shows the best results for the cylindrical PS $\mid P(k_{\perp},k_{\parallel}) \mid$ which has been used to estimate the spherically binned PS $P(k)$ where $k=\sqrt{k_{\perp}^2 + k_{\parallel}^2}$. This has been obtained from the combined nights data for $f=0.6$. The black dashed line shows $[k_{\parallel}]_H$ the predicted boundary of the foreground wedge. We notice that the foreground leakage extends beyond $[k_{\parallel}]_H$. We have selected the entire $k_{\perp}$ range, and $k_{\parallel}$ modes inside the ``21-cm window'', with a buffer region of $0.5-1.0\,\,{\rm Mpc}^{-1}$ outside the wedge boundary as shown in Figure~\ref{fig:pk-sph}. We have included all the modes beyond the green solid lines in the $(k_{\perp},k_{\parallel})$ plane for the spherical binning. Figure \ref{pkbysigma} shows the statistics of the estimated $P(k_{\perp},k_{\parallel})$ for the selected region through a variable `$X$', which is defined by the ratio of the PS to the statistical error due to the system noise at a $(k_{\perp},k_{\parallel})$,
\begin{equation}
X=\frac{P(k_{\perp},k_{\parallel})}{\delta P_{N}(k_{\perp},k_{\parallel})}.
\label{eq:stat}
\end{equation}
We see that the probability density function (P.D.F.) is mostly symmetric within $\mid X \mid\le\,\, 20$ (area demarcated by the vertical black-dashed lines) with mean $mean(X)=\mu=1.21$ and $\sqrt{var(X)}=\sigma_{Est}=6.09$. The negative values of $X$ are consistent with roughly 3 times $\sigma_{Est}$, and no negative values are observed outside $\mid X \mid\le\,\, 20$. This, along with the positive mean within this region, indicate that no negative bias can arise from the modes selected here for the spherical binning. Further, in the scenario where the PS in the selected region are solely dominated by the system noise which follows a Gaussian distribution with zero mean and variance $\sigma_{N}^{2}$, we expect the variable $X$ to follow a Gaussian distribution with $mean(X)=\mu$ and variance $var(X)=\sigma_{Est}^{2}$ as estimated from a sample of N-points, with $\mu$ being comparable to $\frac{\sigma_{Est}}{\sqrt{N}}$ (i.e. $\mu \lesssim \frac{\sigma_{Est}}{\sqrt{N}}$), and as $N \to \infty$, $\sigma_{Est}^{2}\to1$. The standard deviation $\sigma_{Est}>1$ suggests that the estimated r.m.s. statistical fluctuations $\delta P_{N}(k_{\perp},k_{\parallel})$ are underestimated by a factor $\sigma_{Est}$. We note that factors such as RFI, residual point source contributions, residual calibration errors etc., can additionally contribute to the error and may have caused the variance to exceed the predicted value estimated from the system noise only. We account for this underestimation by multiplying the $\delta P_{N}(k_{\perp},k_{\parallel})$ with $\sigma_{Est}$ henceforth to denote the actual error due to statistical fluctuations, $\delta P^{True}_{N}(k_{\perp},k_{\parallel})=\sigma_{Est}\times\delta P_{N}(k_{\perp},k_{\parallel})$. The black-solid curve in Figure \ref{pkbysigma} shows the P.D.F. fitted with a t-distribution. We find that the P.D.F. is well described by the t-distribution within $\mid X \mid\lsim\,\, 20$; however, the long positive tail observed at higher $X$ is not modelled well and is somewhat underestimated by the t-distribution function. From these observations, we interpret that the estimated $P(k_{\perp},k_{\parallel})$ contain contributions from statistical fluctuations as well as residual foreground emission in the region considered here. Note that we have considered $(k_{\perp},k_{\parallel})$ modes up to $\frac{P(k_{\perp},k_{\parallel})}{\delta P^{True}_{N}(k_{\perp},k_{\parallel})}\approx 30$ for the spherical binning.

\begin{figure}
\begin{center}
\includegraphics[width=\columnwidth]{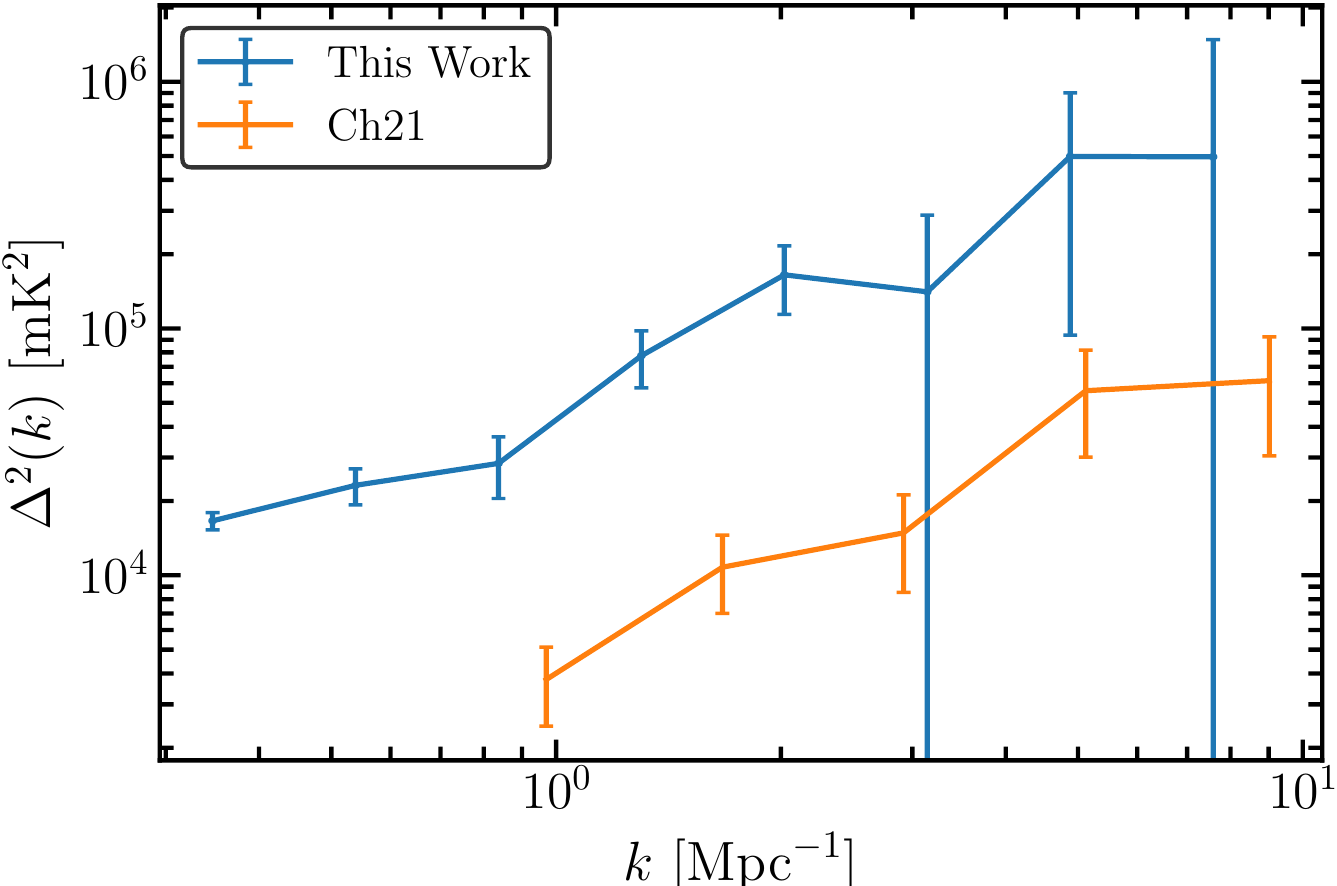}
\caption{The mean square brightness temperature fluctuations $\Delta^2(k)$ shown  as a function of $k$ along with $2 \sigma$ error bars for the selected regions shown in Figure~\ref{fig:pk-sph}. The orange line shows the result at $z=2.19$ from \citetalias{Ch21} along with $2\,\sigma$ error bars as reported in the paper.}
\label{fig:sp-pk}
\end{center}
\end{figure}

We have estimated the spherically binned PS $P(k)$ using the  cylindrical PS $P(k_{\perp},k_{\parallel})$ values in all the $(k_{\perp},k_{\parallel})$ modes which lie beyond the green solid line in Figure~\ref{fig:pk-sph}. The entire $k$ range has been divided into $8$ equally  spaced logarithmic  bins. The solid blue line in Figure~\ref{fig:sp-pk} shows $\mid \Delta^2(k) \mid $ where $\Delta^2(k) \equiv {k^{3}}P(k)/{2\pi^{2}} $ the estimated mean squared brightness temperature fluctuations along with the $2\,\sigma$ error bars. Here $\sigma={k^{3}}[\delta P^{True}_{N}(k)]/{2\pi^{2}}$, where $[\delta P^{True}_{N}(k)]$ is the r.m.s. estimated using the spherically binned PS from $50$ realisations of noise only simulations multiplied by $\sigma_{Est}=6.09$. The values of $ \Delta^2(k)$ and $\sigma$ are tabulated in Table~\ref{tab:ul} for reference. Considering the values of $\Delta^2(k)$, we find that this has the smallest value $\Delta^2(k)=(128.91)^2\,\,{\rm mK}^2$ for the first bin at $k=0.347\,\,{\rm Mpc}^{-1}$. The value of $\Delta^2(k)$ increases approximately as $\Delta^2(k)=(244.16)^2  \, {\rm mK}^2 \, (k/1 {\rm Mpc}^{-1})$ in the entire $k$ range $0.347 < k < 7.584 \,\textrm{Mpc}^{-1}$. The values of $\Delta^2(k)$ in all of the first five bins and the seventh $k$-bin are well in excess of $0 \, +  \, 2 \, \sigma$, and we interpret the values estimated in these bins as arising from residual foregrounds and systematics. The remaining two $k$-bins are consistent with $0 \, +  \, 2 \, \sigma$. We have used the estimated $\Delta^2(k)$ and $\sigma$ values to place $2 \sigma$ upper limits $\Delta_{UL}^{2}(k) = \Delta^{2}(k) + 2\sigma$ on the 21-cm brightness temperature fluctuations at different $k$ bins (Table~\ref{tab:ul}). We find the tightest constraint on the upper limits to be $(133.97)^2 \, {\rm mK}^{2}$ at the smallest bin $k =  0.347\,\textrm{Mpc}^{-1}$. We have used the estimated $\Delta_{UL}^{2}(k)$ to place corresponding $2 \sigma$ upper limits  on $[\Omega_{\rm \HI}b_{\rm \HI}]_{UL}$ (eq.~\ref{eq:ph1}). The values corresponding to the different $k$-bins are tabulated in  Table~\ref{tab:ul}.  We obtain the tightest constraint of $[\Omega_{\rm \HI}b_{\rm \HI}]_{UL} \le  0.23$ from the smallest bin $k = 0.347 \, \textrm{Mpc}^{-1}$.

\begin{table}
\centering
\caption{Spherically binned mean square brightness temperature fluctuations $\Delta^2(k)$ and the corresponding statistical error predictions $\sigma$ for different $k$-bins. The $2\,\sigma$ upper limits  $\Delta^{2}_{UL}(k)=\Delta^{2}(k)+2\,\sigma$ and corresponding $[\Omega_{\rm \HI}b_{\rm \HI}]_{UL}$ values are also provided.}
\begin{tabular}{ccccc}
\hline
\hline
$k$ & $\Delta^2(k)$ & $1\sigma$ & $\Delta_{UL}^{2}(k)$ & $[\Omega_{\rm \HI}b_{\rm \HI}]_{UL}$\\
Mpc$^{-1}$ & (mK)$^2$ & (mK)$^2$ & (mK)$^2$ &\\
\hline
$0.347$ &  $(128.91)^2$ & $(25.79)^2$ & $(133.97)^2$ & $0.230$\\
$0.539$ &  $(152.14)^2$ & $(43.92)^2$ & $(164.33)^2$ & $0.234$\\
$0.837$ &  $(168.54)^2$ & $(62.99)^2$ & $(190.64)^2$ & $0.230$\\
$1.301$ &  $(278.48)^2$ & $(100.37)^2$ & $(312.57)^2$ & $0.326$\\
$2.021$ &  $(406.20)^2$ & $(159.58)^2$ & $(464.68)^2$ & $0.425$\\
$3.141$ &  $(375.19)^2$ & $(271.49)^2$ & $(536.83)^2$ & $0.436$\\
$4.881$ &  $(705.60)^2$ & $(449.38)^2$ & $(949.61)^2$ & $0.694$\\
$7.584$ &  $(704.35)^2$ & $(701.07)^2$ & $(1216.18)^2$ & $0.807$\\
\hline
\label{tab:ul}
\end{tabular}
\end{table}

\citetalias{Ch21} have carried out a multi-redshift analysis of the same observational  data after splitting it into several sub-bands of $8$ MHz each. Their  $z=2.19$  sub-band  is closest  to our analysis,  and we have also shown their results   in Figure~\ref{fig:sp-pk} (orange line). We see that our present analysis extends to substantially smaller $k$ values $(0.347 < k < 7.584 \,\textrm{Mpc}^{-1})$ compared to \citetalias{Ch21} who have considered the $k$-range $1 < k < 10 \,\textrm{Mpc}^{-1}$. We find that in the common $k$-range $1 < k < 8 \,\textrm{Mpc}^{-1}$ our $\Delta^{2}(k)$ estimates are $\sim 7$ times larger than those  of \citetalias{Ch21}. We also see that the $2\,\sigma$ error bars on $\Delta^{2}(k)$  are  smaller in the present  analysis as  compared to \citetalias{Ch21} in the first five $k$-bins. The difference can be attributed to the larger frequency bandwidth $(24.4 \, {\rm MHz})$ used here. For the remaining three $k$-bins, the larger error bars may be attributed to lower sampling at longer baselines. Comparing the $2 \sigma$ upper limits,  \citetalias{Ch21} have obtained $(61.49)^{2} \, {\rm mK}^{2}$ and $0.11$  respectively for $\Delta^{2}(k)$ and $[\Omega_{\rm \HI}b_{\rm \HI}] $ at $k=1 \, {\rm Mpc}^{-1}$  whereas the present analysis reports  $(133.97)^{2} \, {\rm mK}^{2}$ and $0.23$ respectively at $k=0.347 \, {\rm Mpc}^{-1}$.
 
We take  this opportunity to  highlight a key difference between the analysis method of \citetalias{Ch21} and the one used here.   Considering the visibilities measured at the individual baselines, \citetalias{Ch21}  have carried out a Fourier transform along frequency to estimate the visibilities in delay space \citep{Morales04} which were then used \citep{parsons12} to estimate the PS.   The difficulty arises because the missing frequency channels (flagged due to RFI) introduce artefacts in the Fourier transform which corrupt the estimated PS.  \citetalias{Ch21} have overcome this by using  one dimensional complex CLEAN \citep{Parsons_2009}  to compensate  for the missing frequency channels.  In contrast,   the method used here first correlates the visibility data across frequency channels to estimate $C_{\ell}(\Delta \nu)$. There are no missing  frequency separations $\Delta \nu$ 
in the estimated  $C_{\ell}(\Delta \nu)$ even though the visibility data has a substantial number of  missing frequency channels.   We have then used maximum likelihood to estimate the PS which is related to  $C_{\ell}(\Delta \nu)$ through a Fourier transform with respect to $\Delta \nu$.  This method uses only the available frequency channels to estimate the PS, and  it is not necessary   to compensate for the missing frequency channels.  For the present analysis we have validated this using simulations (Section~\ref{sec:simulation}) where the flagging of the simulated data exactly matches that of  the actual data $(55 \%)$. An earlier work \citep{Bh18} has used simulations to demonstrate that the present estimator is able to successfully recover the PS even  when the data in $ 80 \%$ randomly chosen frequency channels are flagged.

\section{Summary and Conclusions}
\label{s7}

\HI 21-cm intensity mapping  is a promising tool to probe the large-scale structures in the Universe across  a wide redshift range.  In this paper we employ the TGE to estimate the MAPS $C_{\ell}(\Delta\nu)$ and the PS  $P(k_{\perp}, k_{\parallel})$  using data from four nights of uGMRT Band 3 observations of the ELAIS-N1 field. Our analysis is restricted to a $24.4\,{\rm MHz}$ sub-band which has a  frequency resolution of $24.4 \,{\rm kHz}$ and is centered at $432.8\, {\rm  MHz}$ which corresponds to 21-cm signal from a  redshift $z=2.28$. 
Compact and discrete sources with  flux densities $> 100\mu  \textrm{Jy}$ within  an area of $1.8 \, {\rm deg^2}$ were identified  and subtracted out.  The residual visibility data  was used for the analysis presented here.  In addition to the individual nights data, we have also analysed the combined  nights data.   

The TGE (eq.~\ref{eq:a4}) uses the measured visibilities to directly estimate $C_{\ell}(\Delta\nu)$ which  characterizes the second order statistics of the sky signal jointly as a function of  the angular multipole $\ell$ and frequency separation $\Delta \nu$. The TGE has three inherent advantages namely (1) it works with the gridded  visibility data which makes it computationally fast; (2) it allows us to taper the sky response which reduces the foreground contamination from bright sources located in the side-lobes and the periphery of the  telescope's field of view; (3) it uses  the data to internally estimate the noise bias  and subtracts this to provide an unbiased estimate of  $C_{\ell}(\Delta\nu)$.  We have used maximum likelihood (eq.~\ref{eq:ML}) to determine  $P(k_{\perp}, k_{\parallel})$  from the estimated $C_{\ell}(\Delta\nu)$, the two being related through a Fourier transform (eq.~\ref{eq:cl_Pk}). 
We have validated the  power-spectrum estimator  using simulations (Section~\ref{sec:simulation}) which  incorporate the flagging, frequency  and   baseline coverage of the actual data. As noted earlier, it is not necessary   to compensate for the missing frequency channels.  Our analysis demonstrates  that  the estimator can recover the  input model  power spectrum with high accuracy over the entire $k$ range used for the  analysis presented in this paper (Figure \ref{fig:pk_val}). 
 
The  May 6 data has the least flagging and the minimum visibility r.m.s. (Table~\ref{tab:flagandrms}), and Figure~\ref{fig:6may_f} shows $C_{\ell}(\Delta\nu)$ for different values of the tapering parameter $f$  considering two values of $\ell$. Note that the tapering is more effective (sky response is narrower) as the value of $f$ is reduced. Figure~\ref{fig:MAPS3D} shows $C_{\ell}(\Delta\nu)$ for the full $(\ell,\Delta \nu)$ range considered here.  Considering $f=5$, which may be loosely interpreted as no tapering, we find that  $C_{\ell}(\Delta\nu)$ exhibits oscillations in $\Delta \nu$, the frequency of the oscillations increases with $\ell$.  We interpret these oscillations as arising from residual compact sources located at large angles from the phase center. We find  that the overall amplitude as well as the amplitude of the oscillations in $C_{\ell}(\Delta\nu)$ both decrease as $f$ is varied from $f=5$ to $f=0.6$. This demonstrates that the TGE is effective in tapering the sky response to suppress the contribution from sources in the outer region of the field of view. Both Figures~\ref{fig:6may_f} and  \ref{fig:MAPS3D} also show the results for the combined nights data with $f=0.6$. Comparing this with May 6 with $f=0.6$, we find that the oscillations and the overall amplitude of $C_{\ell}(\Delta\nu)$ is even further reduced when we consider the combined nights data. This is a direct consequence of the higher   baseline density   (Figure \ref{fig:uv}) which makes the tapering more effective for  the combined nights data  in comparison to the May 6 data. This is due to the fact that tapering in the TGE is implemented through a convolution (eq.~\ref{eq:a1}) which is more effective for the higher baseline density.
 
The different panels of Figure~\ref{fig:Pk_f} shows   $\mid P(k_{\perp}, k_{\parallel}) \mid $ for four  different cases. In all cases,  the large values of $\mid P(k_{\perp}, k_{\parallel}) \mid $  are mainly   localized within the foreground wedge $k_{\parallel} \le  [k_{\parallel}]_{H}$ (horizon), however we also find  some foreground leakage beyond  the predicted wedge boundary. Considering the May 6 data, we find that overall foreground amplitude and also the foreground leakage outside the wedge both come down as the value of $f$ is reduced from $f=5$ to $f=0.6$. There is an even  further reduction when we consider the combined nights data with $f=0.6$.  We see that the combined nights data gives better results in comparison to the May 6 data, and we have used the combined nights data for the subsequent  results and discussion  here.  Considering $f=5.0$ and $f=0.6$,   Figure~\ref{fig:Pk_f_slice} shows  $\mid P(k_{\perp},k_{\parallel}) \mid $ as a function of $k_{\parallel}$ for two different fixed $k_{\perp}$ bins. We find  the largest values of $P(k_{\perp},k_{\parallel})$  $(\sim 10^9 \, {\rm mK}^2 \, {\rm Mpc}^3)$    at $k_{\parallel}=0$. The value of $\mid P(k_{\perp},k_{\parallel}) \mid $ fall  with increasing $k_{\parallel}$, and we find  $\mid P(k_{\perp},k_{\parallel}) \mid  \sim 10^4 \, - \, 10^5  \, {\rm mK}^2 \, {\rm Mpc}^3$  at the largest   $k_{\parallel}$   bins where  the power oscillates between positive and negative values  which are comparable with the $1-\sigma$ error-bars computed from system noise only simulations. We interpret the estimated power in this region as a combination of system noise and some  residual foreground leakage. We also find that the values of  $\mid P(k_{\perp},k_{\parallel}) \mid $ decrease when $f$ is  reduced from $f=5.0$ to $f=0.6$. 
 
An interesting feature seen in the various panels of Figure~\ref{fig:Pk_f}  and also in both the panels of Figure~\ref{fig:Pk_f_slice} is that the value of $P(k_{\perp},k_{\parallel})$ does not decrease monotonically with increasing $k_{\parallel} $. Rather, it   initially decreases  and then increases again  just beyond the horizon  $k_{\parallel} \approx [k_{\parallel}]_{H}$ after which it decreases again. This is more clearly visible at large $k_{\perp}$.  We identify the rise in $P(k_{\perp},k_{\parallel})$ close to  the horizon limit as the pitchfork effect  which has been reported earlier \citep{thyag15,kohn16,gehlot17} in low frequency  observations $(\sim 150 \, {\rm MHz})$  which target the EoR 21-cm signal.   The present work is possibly the first time that this effect is being reported  at higher frequencies which target the post-reionization 21-cm signal. 
 
The solid green  curve in Figure~\ref{fig:pk-sph} demarcates the region of $(k_{\perp},k_{\parallel})$ space which we have identified to be relatively foreground-free and has been used to estimate the spherically binned power spectrum $P(k)$. We use the variable $X$, defined in eq.~(\ref{eq:stat}), to study the statistics of the estimated PS in this region (Figure \ref{pkbysigma}). We find that for $\mid X\mid\lsim\,\,20$ the P.D.F. is roughly symmetric with a positive mean $\mu=1.21$. For $\mid X\mid\lsim\,\,20$, the P.D.F. is well described the t-distribution, beyond which $(X>20)$ the t-distribution function underestimates the P.D.F. This indicates that the PS consists of some noise contributions as well as residual foregrounds. Modes within $\mid X\mid\lsim\,\,20$ have a standard deviation $\sigma_{Est}=6.09$, which suggests that the r.m.s. fluctuations estimated using the noise simulations underestimate the true errors by a factor $\sigma_{Est}=6.09$. We rectify for this by considering the true errors as $\delta P^{True}_{N}(k_{\perp},k_{\parallel})=\sigma_{Est}\times\delta P_{N}(k_{\perp},k_{\parallel})$, which we carry forward for further analysis. Figure~\ref{fig:sp-pk} shows the mean square brightness temperature fluctuations $\Delta^2(k)$  along with $2 \sigma$ error bars  considering $8$ bins across the  range $0.347 \le k \le  7.584 \,\textrm{Mpc}^{-1}$. Table \ref{tab:ul} lists these values along with $\Delta_{UL}^{2}(k)$ the corresponding $2 \sigma$ upper limits. We find the tightest $2 \sigma$ upper limit of $\Delta_{UL}^{2}(k)\le(133.97)^2\,{\rm mK}^{2}$ at $k=0.347\,{\rm Mpc}^{-1}$ which translates to an upper limit  $[\Omega_{\rm \HI}b_{\rm \HI}]_{UL}\le 0.23$. \citetalias{Ch21} reported $\Delta_{UL}^{2}(k)\le(61.49)^{2} \, {\rm mK}^{2}$ and $[\Omega_{\rm \HI} b_{\rm \HI}]_{UL} \le 0.11$ at $k=1 \, {\rm Mpc}^{-1}$. The upper limits presented here are still orders of magnitude larger than the expected signal corresponding to $\Omega_{\rm \HI} \sim 10^{-3}$ and $b_{\rm \HI} \sim 2 $.

\section*{Acknowledgements}
We thank the scientific editor and the anonymous referee for their comments which have helped us in improving this work. We thank the staff of GMRT for making this observation possible. GMRT is run by National Centre for Radio Astrophysics of the Tata Institute of Fundamental Research. AG would like to acknowledge IUCAA, Pune for providing support through the associateship programme. SB would like to acknowledge funding provided under the MATRICS grant SERB/F/9805/2019-2020 of the Science \& Engineering Research Board, a statutory body of Department of Science \& Technology (DST), Government of India. Part of this work has used the Supercomputing facility `PARAM Shakti' of IIT Kharagpur established under National Supercomputing Mission (NSM), Government of India and supported by Centre for Development of Advanced Computing (CDAC), Pune.

%%%%%%%%%%%%%%%%%%%%%%%%%%%%%%%%%%%%%%%%%%%%%%%%%%
\section*{Data Availability}

The data from this study are available upon reasonable request to the corresponding author.


\bibliographystyle{mnras}
\bibliography{myref}

\begin{thebibliography}{}
\makeatletter
\relax
\def\mn@urlcharsother{\let\do\@makeother \do\$\do\&\do\#\do\^\do\_\do\%\do\~}
\def\mn@doi{\begingroup\mn@urlcharsother \@ifnextchar [ {\mn@doi@}
  {\mn@doi@[]}}
\def\mn@doi@[#1]#2{\def\@tempa{#1}\ifx\@tempa\@empty \href
  {http://dx.doi.org/#2} {doi:#2}\else \href {http://dx.doi.org/#2} {#1}\fi
  \endgroup}
\def\mn@eprint#1#2{\mn@eprint@#1:#2::\@nil}
\def\mn@eprint@arXiv#1{\href {http://arxiv.org/abs/#1} {{\tt arXiv:#1}}}
\def\mn@eprint@dblp#1{\href {http://dblp.uni-trier.de/rec/bibtex/#1.xml}
  {dblp:#1}}
\def\mn@eprint@#1:#2:#3:#4\@nil{\def\@tempa {#1}\def\@tempb {#2}\def\@tempc
  {#3}\ifx \@tempc \@empty \let \@tempc \@tempb \let \@tempb \@tempa \fi \ifx
  \@tempb \@empty \def\@tempb {arXiv}\fi \@ifundefined
  {mn@eprint@\@tempb}{\@tempb:\@tempc}{\expandafter \expandafter \csname
  mn@eprint@\@tempb\endcsname \expandafter{\@tempc}}}

\bibitem[\protect\citeauthoryear{{Ali} \& {Bharadwaj}}{{Ali} \&
  {Bharadwaj}}{2014}]{ali14}
{Ali} S.~S.,  {Bharadwaj} S.,  2014, \mn@doi [\japa]
  {10.1007/s12036-014-9301-1}, \href
  {https://ui.adsabs.harvard.edu/abs/2014JApA...35..157A} {35, 157}

\bibitem[\protect\citeauthoryear{Anderson et~al.,}{Anderson
  et~al.}{2018}]{Anderson2018}
Anderson C.~J.,  et~al., 2018, \mn@doi [Monthly Notices of the Royal
  Astronomical Society] {10.1093/mnras/sty346}, 476, 3382

\bibitem[\protect\citeauthoryear{{Bharadwaj} \& {Ali}}{{Bharadwaj} \&
  {Ali}}{2005}]{BA5}
{Bharadwaj} S.,  {Ali} S.~S.,  2005, \mn@doi [\mnras]
  {10.1111/j.1365-2966.2004.08604.x}, \href
  {http://adsabs.harvard.edu/abs/2005MNRAS.356.1519B} {356, 1519}

\bibitem[\protect\citeauthoryear{{Bharadwaj} \& {Pandey}}{{Bharadwaj} \&
  {Pandey}}{2003}]{Bharadwaj-Pandey-2003}
{Bharadwaj} S.,  {Pandey} S.~K.,  2003, \mn@doi [\japa] {10.1007/BF03012189},
  \href {https://ui.adsabs.harvard.edu/abs/2003JApA...24...23B} {24, 23}

\bibitem[\protect\citeauthoryear{{Bharadwaj} \& {Sethi}}{{Bharadwaj} \&
  {Sethi}}{2001}]{BS01}
{Bharadwaj} S.,  {Sethi} S.~K.,  2001, \mn@doi [\japa] {10.1007/BF02702273},
  \href {http://adsabs.harvard.edu/abs/2001JApA...22..293B} {22, 293}

\bibitem[\protect\citeauthoryear{{Bharadwaj} \& {Srikant}}{{Bharadwaj} \&
  {Srikant}}{2004}]{bh_sri2004}
{Bharadwaj} S.,  {Srikant} P.~S.,  2004, \mn@doi [\japa] {10.1007/BF02702289},
  \href {https://ui.adsabs.harvard.edu/abs/2004JApA...25...67B} {25, 67}

\bibitem[\protect\citeauthoryear{{Bharadwaj}, {Nath}  \& {Sethi}}{{Bharadwaj}
  et~al.}{2001}]{BNS}
{Bharadwaj} S.,  {Nath} B.~B.,   {Sethi} S.~K.,  2001, \mn@doi [\japa]
  {10.1007/BF02933588}, \href
  {https://ui.adsabs.harvard.edu/abs/2001JApA...22...21B} {22, 21}

\bibitem[\protect\citeauthoryear{Bharadwaj, Sethi  \& Saini}{Bharadwaj
  et~al.}{2009}]{Bh09}
Bharadwaj S.,  Sethi S.~K.,   Saini T.~D.,  2009, \mn@doi [Phys. Rev. D]
  {10.1103/PhysRevD.79.083538}, 79, 083538

\bibitem[\protect\citeauthoryear{Bharadwaj, Pal, Choudhuri  \& Dutta}{Bharadwaj
  et~al.}{2018}]{Bh18}
Bharadwaj S.,  Pal S.,  Choudhuri S.,   Dutta P.,  2018, \mn@doi [\mnras]
  {10.1093/mnras/sty3501}, 483, 5694

\bibitem[\protect\citeauthoryear{{Bowman}, {Morales}  \& {Hewitt}}{{Bowman}
  et~al.}{2009}]{bowman09}
{Bowman} J.~D.,  {Morales} M.~F.,   {Hewitt} J.~N.,  2009, \mn@doi [\apj]
  {10.1088/0004-637X/695/1/183}, \href
  {http://adsabs.harvard.edu/abs/2009ApJ...695..183B} {695, 183}

\bibitem[\protect\citeauthoryear{{Bull}, {Camera}, {Raccanelli}, {Blake},
  {Ferreira}, {Santos}  \& {Schwarz}}{{Bull} et~al.}{2015}]{SKA15}
{Bull} P.,  {Camera} S.,  {Raccanelli} A.,  {Blake} C.,  {Ferreira} P.,
  {Santos} M.,   {Schwarz} D.~J.,  2015, in AASKA14. p.~24 (\mn@eprint {arXiv}
  {1501.04088})

\bibitem[\protect\citeauthoryear{{CHIME Collaboration} et~al.,}{{CHIME
  Collaboration} et~al.}{2022}]{chime22}
{CHIME Collaboration} et~al., 2022, arXiv e-prints, \href
  {https://ui.adsabs.harvard.edu/abs/2022arXiv220201242C} {p. arXiv:2202.01242}

\bibitem[\protect\citeauthoryear{Chakraborty et~al.,}{Chakraborty
  et~al.}{2019a}]{Cha1}
Chakraborty A.,  et~al., 2019a, \mn@doi [\mnras] {10.1093/mnras/stz1580}, 487,
  4102

\bibitem[\protect\citeauthoryear{Chakraborty et~al.,}{Chakraborty
  et~al.}{2019b}]{Cha2}
Chakraborty A.,  et~al., 2019b, \mn@doi [\mnras] {10.1093/mnras/stz2533}, 490,
  243

\bibitem[\protect\citeauthoryear{Chakraborty et~al.,}{Chakraborty
  et~al.}{2021}]{Ch21}
Chakraborty A.,  et~al., 2021, \mn@doi [\apjl] {10.3847/2041-8213/abd17a}, 907,
  L7

\bibitem[\protect\citeauthoryear{Chang, Pen, Peterson  \& McDonald}{Chang
  et~al.}{2008}]{Chang08}
Chang T.-C.,  Pen U.-L.,  Peterson J.~B.,   McDonald P.,  2008, \mn@doi [Phys.
  Rev. Lett.] {10.1103/PhysRevLett.100.091303}, 100, 091303

\bibitem[\protect\citeauthoryear{{Chang}, {Pen}, {Bandura}  \&
  {Peterson}}{{Chang} et~al.}{2010}]{chang10}
{Chang} T.-C.,  {Pen} U.-L.,  {Bandura} K.,   {Peterson} J.~B.,  2010, \mn@doi
  [\nat] {10.1038/nature09187}, \href
  {https://ui.adsabs.harvard.edu/abs/2010Natur.466..463C} {466, 463}

\bibitem[\protect\citeauthoryear{Chen}{Chen}{2012}]{tian}
Chen X.,  2012, \mn@doi [International Journal of Modern Physics: Conference
  Series] {10.1142/S2010194512006459}, 12, 256

\bibitem[\protect\citeauthoryear{Choudhuri, Bharadwaj, Ghosh  \& Ali}{Choudhuri
  et~al.}{2014}]{samir14}
Choudhuri S.,  Bharadwaj S.,  Ghosh A.,   Ali S.~S.,  2014, \mn@doi [\mnras]
  {10.1093/mnras/stu2027}, 445, 4351

\bibitem[\protect\citeauthoryear{Choudhuri, Bharadwaj, Roy, Ghosh  \&
  Ali}{Choudhuri et~al.}{2016a}]{samir16}
Choudhuri S.,  Bharadwaj S.,  Roy N.,  Ghosh A.,   Ali S.~S.,  2016a, \mn@doi
  [\mnras] {10.1093/mnras/stw607}, 459, 151

\bibitem[\protect\citeauthoryear{Choudhuri, Bharadwaj, Chatterjee, Ali, Roy  \&
  Ghosh}{Choudhuri et~al.}{2016b}]{samir17}
Choudhuri S.,  Bharadwaj S.,  Chatterjee S.,  Ali S.~S.,  Roy N.,   Ghosh A.,
  2016b, \mn@doi [\mnras] {10.1093/mnras/stw2254}, 463, 4093

\bibitem[\protect\citeauthoryear{Choudhuri, Bharadwaj, Ali, Roy, Intema  \&
  Ghosh}{Choudhuri et~al.}{2017}]{samir17a}
Choudhuri S.,  Bharadwaj S.,  Ali S.~S.,  Roy N.,  Intema H.~T.,   Ghosh A.,
  2017, \mn@doi [\mnras: Letters] {10.1093/mnrasl/slx066}, 470, L11

\bibitem[\protect\citeauthoryear{Choudhuri, Ghosh, Roy, Bharadwaj, Intema  \&
  Ali}{Choudhuri et~al.}{2020}]{samir20}
Choudhuri S.,  Ghosh A.,  Roy N.,  Bharadwaj S.,  Intema H.~T.,   Ali S.~S.,
  2020, \mn@doi [\mnras] {10.1093/mnras/staa762}, 494, 1936

\bibitem[\protect\citeauthoryear{Datta, Choudhury  \& Bharadwaj}{Datta
  et~al.}{2007}]{KD07}
Datta K.~K.,  Choudhury T.~R.,   Bharadwaj S.,  2007, \mn@doi [\mnras]
  {10.1111/j.1365-2966.2007.11747.x}, 378, 119

\bibitem[\protect\citeauthoryear{{Datta}, {Bowman}  \& {Carilli}}{{Datta}
  et~al.}{2010}]{adatta10}
{Datta} A.,  {Bowman} J.~D.,   {Carilli} C.~L.,  2010, \mn@doi [\apj]
  {10.1088/0004-637X/724/1/526}, \href
  {https://ui.adsabs.harvard.edu/abs/2010ApJ...724..526D} {724, 526}

\bibitem[\protect\citeauthoryear{Eisenstein \& Hu}{Eisenstein \&
  Hu}{1998}]{Eisenstein_1998}
Eisenstein D.~J.,  Hu W.,  1998, \mn@doi [\apj] {10.1086/305424}, 496, 605

\bibitem[\protect\citeauthoryear{{Gehlot} et~al.,}{{Gehlot}
  et~al.}{2018}]{gehlot17}
{Gehlot} B.~K.,  et~al., 2018, \mn@doi [\mnras] {10.1093/mnras/sty1095}, \href
  {https://ui.adsabs.harvard.edu/abs/2018MNRAS.478.1484G} {478, 1484}

\bibitem[\protect\citeauthoryear{Ghosh, Bharadwaj, Ali  \& Chengalur}{Ghosh
  et~al.}{2011a}]{ghosh1}
Ghosh A.,  Bharadwaj S.,  Ali S.~S.,   Chengalur J.~N.,  2011a, \mn@doi
  [\mnras] {10.1111/j.1365-2966.2010.17853.x}, 411, 2426

\bibitem[\protect\citeauthoryear{Ghosh, Bharadwaj, Ali  \& Chengalur}{Ghosh
  et~al.}{2011b}]{ghosh2}
Ghosh A.,  Bharadwaj S.,  Ali S.~S.,   Chengalur J.~N.,  2011b, \mn@doi
  [\mnras] {10.1111/j.1365-2966.2011.19649.x}, 418, 2584

\bibitem[\protect\citeauthoryear{Ghosh, Prasad, Bharadwaj, Ali  \&
  Chengalur}{Ghosh et~al.}{2012}]{ghosh3}
Ghosh A.,  Prasad J.,  Bharadwaj S.,  Ali S.~S.,   Chengalur J.~N.,  2012,
  \mn@doi [\mnras] {10.1111/j.1365-2966.2012.21889.x}, 426, 3295

\bibitem[\protect\citeauthoryear{{Gupta} et~al.,}{{Gupta} et~al.}{2017}]{uGMRT}
{Gupta} Y.,  et~al., 2017, Current Science, \href
  {https://ui.adsabs.harvard.edu/abs/2017CSci..113..707G} {113, 707}

\bibitem[\protect\citeauthoryear{{Hazra} \& {Guha Sarkar}}{{Hazra} \& {Guha
  Sarkar}}{2012}]{Hazra2012}
{Hazra} D.~K.,  {Guha Sarkar} T.,  2012, \mn@doi [\prl]
  {10.1103/PhysRevLett.109.121301}, \href
  {https://ui.adsabs.harvard.edu/abs/2012PhRvL.109l1301H} {109, 121301}

\bibitem[\protect\citeauthoryear{{Kennedy} \& {Bull}}{{Kennedy} \&
  {Bull}}{2021}]{Kennedy21}
{Kennedy} F.,  {Bull} P.,  2021, \mn@doi [\mnras] {10.1093/mnras/stab1814},
  \href {https://ui.adsabs.harvard.edu/abs/2021MNRAS.506.2638K} {506, 2638}

\bibitem[\protect\citeauthoryear{{Kohn} et~al.,}{{Kohn} et~al.}{2016}]{kohn16}
{Kohn} S.~A.,  et~al., 2016, \mn@doi [\apj] {10.3847/0004-637X/823/2/88}, \href
  {https://ui.adsabs.harvard.edu/abs/2016ApJ...823...88K} {823, 88}

\bibitem[\protect\citeauthoryear{{Kumar}, {Dutta}  \& {Roy}}{{Kumar}
  et~al.}{2020}]{JK20}
{Kumar} J.,  {Dutta} P.,   {Roy} N.,  2020, \mn@doi [\mnras]
  {10.1093/mnras/staa1371}, \href
  {https://ui.adsabs.harvard.edu/abs/2020MNRAS.495.3683K} {495, 3683}

\bibitem[\protect\citeauthoryear{{Kumar}, {Dutta}, {Choudhuri}  \&
  {Roy}}{{Kumar} et~al.}{2022}]{JK22}
{Kumar} J.,  {Dutta} P.,  {Choudhuri} S.,   {Roy} N.,  2022, \mn@doi [\mnras]
  {10.1093/mnras/stac499}, \href
  {https://ui.adsabs.harvard.edu/abs/2022MNRAS.512..186K} {512, 186}

\bibitem[\protect\citeauthoryear{{Masui} et~al.,}{{Masui}
  et~al.}{2013}]{masui2013}
{Masui} K.~W.,  et~al., 2013, \mn@doi [\apjl] {10.1088/2041-8205/763/1/L20},
  \href {https://ui.adsabs.harvard.edu/abs/2013ApJ...763L..20M} {763, L20}

\bibitem[\protect\citeauthoryear{Mazumder, Chakraborty, Datta, Choudhuri, Roy,
  Wadadekar  \& Ishwara-Chandra}{Mazumder et~al.}{2020}]{M20}
Mazumder A.,  Chakraborty A.,  Datta A.,  Choudhuri S.,  Roy N.,  Wadadekar Y.,
    Ishwara-Chandra C.~H.,  2020, \mn@doi [\mnras] {10.1093/mnras/staa1317},
  495, 4071

\bibitem[\protect\citeauthoryear{{McMullin}, {Waters}, {Schiebel}, {Young}  \&
  {Golap}}{{McMullin} et~al.}{2007}]{casa07}
{McMullin} J.~P.,  {Waters} B.,  {Schiebel} D.,  {Young} W.,   {Golap} K.,
  2007, in {Shaw} R.~A.,  {Hill} F.,   {Bell} D.~J.,  eds,  Astronomical
  Society of the Pacific Conference Series Vol. 376, Astronomical Data Analysis
  Software and Systems XVI. p.~127

\bibitem[\protect\citeauthoryear{{Mondal}, {Bharadwaj}, {Iliev}, {Datta},
  {Majumdar}, {Shaw}  \& {Sarkar}}{{Mondal} et~al.}{2019}]{Mondal19}
{Mondal} R.,  {Bharadwaj} S.,  {Iliev} I.~T.,  {Datta} K.~K.,  {Majumdar} S.,
  {Shaw} A.~K.,   {Sarkar} A.~K.,  2019, \mn@doi [\mnras]
  {10.1093/mnrasl/sly226}, \href
  {https://ui.adsabs.harvard.edu/abs/2019MNRAS.483L.109M} {483, L109}

\bibitem[\protect\citeauthoryear{{Morales} \& {Hewitt}}{{Morales} \&
  {Hewitt}}{2004}]{Morales04}
{Morales} M.~F.,  {Hewitt} J.,  2004, \mn@doi [\apj] {10.1086/424437}, \href
  {https://ui.adsabs.harvard.edu/abs/2004ApJ...615....7M} {615, 7}

\bibitem[\protect\citeauthoryear{Morales, Hazelton, Sullivan  \&
  Beardsley}{Morales et~al.}{2012}]{Morales_2012}
Morales M.~F.,  Hazelton B.,  Sullivan I.,   Beardsley A.,  2012, \mn@doi
  [\apj] {10.1088/0004-637x/752/2/137}, 752, 137

\bibitem[\protect\citeauthoryear{Murray \& Trott}{Murray \&
  Trott}{2018}]{Murray_2018}
Murray S.~G.,  Trott C.~M.,  2018, \mn@doi [\apj] {10.3847/1538-4357/aaebfa},
  869, 25

\bibitem[\protect\citeauthoryear{{Newburgh} et~al.,}{{Newburgh}
  et~al.}{2016}]{Newburgh16}
{Newburgh} L.~B.,  et~al., 2016, in {Hall} H.~J.,  {Gilmozzi} R.,   {Marshall}
  H.~K.,  eds,  Society of Photo-Optical Instrumentation Engineers (SPIE)
  Conference Series Vol. 9906, Ground-based and Airborne Telescopes VI. p.
  99065X (\mn@eprint {arXiv} {1607.02059}), \mn@doi{10.1117/12.2234286}

\bibitem[\protect\citeauthoryear{{Noterdaeme} et~al.,}{{Noterdaeme}
  et~al.}{2012}]{Not}
{Noterdaeme} P.,  et~al., 2012, \mn@doi [\aap] {10.1051/0004-6361/201220259},
  \href {https://ui.adsabs.harvard.edu/abs/2012A&A...547L...1N} {547, L1}

\bibitem[\protect\citeauthoryear{{Nuttall}}{{Nuttall}}{1981}]{nut81}
{Nuttall} A.~H.,  1981, IEEE Transactions on Acoustics Speech and Signal
  Processing, \href {https://ui.adsabs.harvard.edu/abs/1981ITASS..29...84N}
  {29, 84}

\bibitem[\protect\citeauthoryear{{Offringa, A. R.}, {van de Gronde, J. J.}  \&
  {Roerdink, J. B. T. M.}}{{Offringa, A. R.} et~al.}{2012}]{Off12}
{Offringa, A. R.} {van de Gronde, J. J.}  {Roerdink, J. B. T. M.} 2012, \mn@doi
  [A\&A] {10.1051/0004-6361/201118497}, 539, A95

\bibitem[\protect\citeauthoryear{Offringa, de Bruyn, Biehl, Zaroubi, Bernardi
  \& Pandey}{Offringa et~al.}{2010}]{Off10}
Offringa A.~R.,  de Bruyn A.~G.,  Biehl M.,  Zaroubi S.,  Bernardi G.,   Pandey
  V.~N.,  2010, \mn@doi [\mnras] {10.1111/j.1365-2966.2010.16471.x}, 405, 155

\bibitem[\protect\citeauthoryear{{Pal}, {Bharadwaj}, {Ghosh}  \&
  {Choudhuri}}{{Pal} et~al.}{2021}]{Pal20}
{Pal} S.,  {Bharadwaj} S.,  {Ghosh} A.,   {Choudhuri} S.,  2021, \mn@doi
  [\mnras] {10.1093/mnras/staa3831}, \href
  {https://ui.adsabs.harvard.edu/abs/2021MNRAS.501.3378P} {501, 3378}

\bibitem[\protect\citeauthoryear{{Parsons} \& {Backer}}{{Parsons} \&
  {Backer}}{2009}]{Parsons_2009}
{Parsons} A.~R.,  {Backer} D.~C.,  2009, \mn@doi [\aj]
  {10.1088/0004-6256/138/1/219}, \href
  {https://ui.adsabs.harvard.edu/abs/2009AJ....138..219P} {138, 219}

\bibitem[\protect\citeauthoryear{Parsons, Pober, Aguirre, Carilli, Jacobs  \&
  Moore}{Parsons et~al.}{2012}]{parsons12}
Parsons A.~R.,  Pober J.~C.,  Aguirre J.~E.,  Carilli C.~L.,  Jacobs D.~C.,
  Moore D.~F.,  2012, \mn@doi [\apj] {10.1088/0004-637x/756/2/165}, 756, 165

\bibitem[\protect\citeauthoryear{Pen, Staveley-Smith, Peterson  \& Chang}{Pen
  et~al.}{2009}]{Pen2009a}
Pen U.-L.,  Staveley-Smith L.,  Peterson J.~B.,   Chang T.-C.,  2009, \mn@doi
  [\mnras: Letters] {10.1111/j.1745-3933.2008.00581.x}, 394, L6

\bibitem[\protect\citeauthoryear{{Planck Collaboration} et~al.,}{{Planck
  Collaboration} et~al.}{2020}]{Planck18f}
{Planck Collaboration} et~al., 2020, \mn@doi [\aap]
  {10.1051/0004-6361/201833910}, \href
  {https://ui.adsabs.harvard.edu/abs/2020A&A...641A...6P} {641, A6}

\bibitem[\protect\citeauthoryear{{Pober} et~al.,}{{Pober}
  et~al.}{2016}]{pober16}
{Pober} J.~C.,  et~al., 2016, \mn@doi [\apj] {10.3847/0004-637X/819/1/8}, \href
  {https://ui.adsabs.harvard.edu/abs/2016ApJ...819....8P} {819, 8}

\bibitem[\protect\citeauthoryear{{Saha}, {Bharadwaj}, {Roy}, {Choudhuri}  \&
  {Chattopadhyay}}{{Saha} et~al.}{2019}]{Preetha19}
{Saha} P.,  {Bharadwaj} S.,  {Roy} N.,  {Choudhuri} S.,   {Chattopadhyay} D.,
  2019, \mn@doi [\mnras] {10.1093/mnras/stz2528}, \href
  {https://ui.adsabs.harvard.edu/abs/2019MNRAS.489.5866S} {489, 5866}

\bibitem[\protect\citeauthoryear{Saha, Bharadwaj, Chakravorty, Roy, Choudhuri,
  Günther  \& Smith}{Saha et~al.}{2021}]{Preetha21}
Saha P.,  Bharadwaj S.,  Chakravorty S.,  Roy N.,  Choudhuri S.,  Günther
  H.~M.,   Smith R.~K.,  2021, \mn@doi [\mnras] {10.1093/mnras/stab446}, 502,
  5313

\bibitem[\protect\citeauthoryear{Sarkar, Bharadwaj  \& Anathpindika}{Sarkar
  et~al.}{2016}]{Deb16}
Sarkar D.,  Bharadwaj S.,   Anathpindika S.,  2016, \mn@doi [\mnras]
  {10.1093/mnras/stw1111}, 460, 4310

\bibitem[\protect\citeauthoryear{Seo, Dodelson, Marriner, Mcginnis, Stebbins,
  Stoughton  \& Vallinotto}{Seo et~al.}{2010}]{Seo_2010}
Seo H.-J.,  Dodelson S.,  Marriner J.,  Mcginnis D.,  Stebbins A.,  Stoughton
  C.,   Vallinotto A.,  2010, \mn@doi [\apj] {10.1088/0004-637x/721/1/164},
  721, 164

\bibitem[\protect\citeauthoryear{{Subrahmanya}, {Manoharan}  \&
  {Chengalur}}{{Subrahmanya} et~al.}{2017}]{OWFA}
{Subrahmanya} C.~R.,  {Manoharan} P.~K.,   {Chengalur} J.~N.,  2017, \mn@doi
  [\japa] {10.1007/s12036-017-9430-4}, \href
  {https://ui.adsabs.harvard.edu/abs/2017JApA...38...10S} {38, 10}

\bibitem[\protect\citeauthoryear{{Swarup} et~al.,}{{Swarup}
  et~al.}{1971}]{GS71}
{Swarup} G.,  et~al., 1971, \mn@doi [Nature Physical Science]
  {10.1038/physci230185a0}, \href
  {https://ui.adsabs.harvard.edu/abs/1971NPhS..230..185S} {230, 185}

\bibitem[\protect\citeauthoryear{{Swarup}, {Ananthakrishnan}, {Kapahi}, {Rao},
  {Subrahmanya}  \& {Kulkarni}}{{Swarup} et~al.}{1991}]{swarup91}
{Swarup} G.,  {Ananthakrishnan} S.,  {Kapahi} V.~K.,  {Rao} A.~P.,
  {Subrahmanya} C.~R.,   {Kulkarni} V.~K.,  1991, Current Science, \href
  {https://ui.adsabs.harvard.edu/abs/1991CuSc...60...95S} {60, 95}

\bibitem[\protect\citeauthoryear{Switzer et~al.,}{Switzer et~al.}{2013}]{SW13}
Switzer E.~R.,  et~al., 2013, \mn@doi [\mnras: Letters]
  {10.1093/mnrasl/slt074}, 434, L46

\bibitem[\protect\citeauthoryear{{Thyagarajan} et~al.,}{{Thyagarajan}
  et~al.}{2015a}]{thyag15_1}
{Thyagarajan} N.,  et~al., 2015a, \mn@doi [\apj] {10.1088/0004-637X/804/1/14},
  \href {https://ui.adsabs.harvard.edu/abs/2015ApJ...804...14T} {804, 14}

\bibitem[\protect\citeauthoryear{{Thyagarajan} et~al.,}{{Thyagarajan}
  et~al.}{2015b}]{thyag15}
{Thyagarajan} N.,  et~al., 2015b, \mn@doi [\apjl]
  {10.1088/2041-8205/807/2/L28}, \href
  {https://ui.adsabs.harvard.edu/abs/2015ApJ...807L..28T} {807, L28}

\bibitem[\protect\citeauthoryear{{Thyagarajan}, {Parsons}, {DeBoer}, {Bowman},
  {Ewall-Wice}, {Neben}  \& {Patra}}{{Thyagarajan}
  et~al.}{2016}]{Thyagarajan_2016}
{Thyagarajan} N.,  {Parsons} A.~R.,  {DeBoer} D.~R.,  {Bowman} J.~D.,
  {Ewall-Wice} A.~M.,  {Neben} A.~R.,   {Patra} N.,  2016, \mn@doi [\apj]
  {10.3847/0004-637X/825/1/9}, \href
  {https://ui.adsabs.harvard.edu/abs/2016ApJ...825....9T} {825, 9}

\bibitem[\protect\citeauthoryear{{Vedantham}, {Udaya Shankar}  \&
  {Subrahmanyan}}{{Vedantham} et~al.}{2012}]{vedantham12}
{Vedantham} H.,  {Udaya Shankar} N.,   {Subrahmanyan} R.,  2012, \mn@doi [\apj]
  {10.1088/0004-637X/745/2/176}, \href
  {https://ui.adsabs.harvard.edu/abs/2012ApJ...745..176V} {745, 176}

\bibitem[\protect\citeauthoryear{Visbal, Loeb  \& Wyithe}{Visbal
  et~al.}{2009}]{Visbal_2009}
Visbal E.,  Loeb A.,   Wyithe S.,  2009, \mn@doi [\jcap]
  {10.1088/1475-7516/2009/10/030}, 2009, 030

\bibitem[\protect\citeauthoryear{Wolz et~al.,}{Wolz et~al.}{2021}]{Wolz2021}
Wolz L.,  et~al., 2021, \mn@doi [Monthly Notices of the Royal Astronomical
  Society] {10.1093/mnras/stab3621}, 510, 3495

\bibitem[\protect\citeauthoryear{Wuensche}{Wuensche}{2019}]{Wuensche_2019}
Wuensche C.,  2019, \mn@doi [Journal of Physics: Conference Series]
  {10.1088/1742-6596/1269/1/012002}, 1269, 012002

\bibitem[\protect\citeauthoryear{Wyithe, Loeb  \& Geil}{Wyithe
  et~al.}{2008}]{w08}
Wyithe J. S.~B.,  Loeb A.,   Geil P.~M.,  2008, \mn@doi [\mnras]
  {10.1111/j.1365-2966.2007.12631.x}, 383, 1195

\bibitem[\protect\citeauthoryear{{Zafar}, {P{\'e}roux}, {Popping}, {Milliard},
  {Deharveng}  \& {Frank}}{{Zafar} et~al.}{2013}]{Zafar}
{Zafar} T.,  {P{\'e}roux} C.,  {Popping} A.,  {Milliard} B.,  {Deharveng}
  J.~M.,   {Frank} S.,  2013, \mn@doi [\aap] {10.1051/0004-6361/201321154},
  \href {https://ui.adsabs.harvard.edu/abs/2013A&A...556A.141Z} {556, A141}

\makeatother
\end{thebibliography}

%%%%%%%%%%%%%%%%%%%%%%%%%%%%%%%%%%%%%%%%%%%%%%%%%%

% Don't change these lines
\bsp	% typesetting comment
\label{lastpage}
\end{document}